\documentclass[useAMS,usenatbib]{mn2e}
\usepackage{natbib, aas_macros}

\usepackage{graphicx, color, url}

\newcommand {\apgt} {\ {\raise-.5ex\hbox{$\buildrel>\over\sim$}}\ }
\newcommand {\aplt} {\ {\raise-.5ex\hbox{$\buildrel<\over\sim$}}\ }

\title[Cluster initial mass function]{The initial mass function of
  star clusters that form in turbulent molecular clouds}

\author[M. S. Fujii and S. Portegies Zwart]
{M. S. Fujii$^{1}$
\thanks{E-mail: michiko.fujii@nao.ac.jp(MSF); spz@strw.leidenuniv.nl(SPZ)} 
and S. Portegies Zwart$^{2}$\footnotemark[1]\\
$^{1}$Division of Theoretical Astronomy, National Astronomical Observatory of Japan 2-21-1 Osawa, Mitaka, Tokyo 181-8588, Japan\\
$^{2}$Leiden Observatory, Leiden University, NL-2300RA Leiden, The Netherlands}
\begin{document}

\date{Accepted . Received ; in original form 1988 October 11}

\pagerange{\pageref{firstpage}--\pageref{lastpage}} \pubyear{2002}

\maketitle

\label{firstpage}

\begin{abstract}

We simulate the formation and evolution of young star clusters using
the combination of smoothed-particle hydrodynamical (SPH) simulations
and direct $N$-body simulations. We start by performing SPH
simulations of the giant molecular cloud with a turbulent velocity
field, a mass of $4\times 10^4$ to $5\times 10^6M_{\odot}$, and a
density between $\rho \sim 1.7\times 10^3$ and $170 {\rm cm}^{-3}$.
We continue the hydrodynamical simulations for a free-fall time scale
($t_{\rm ff} \simeq 0.83$\,Myr and 2.5 Myr), and analyze the resulting
structure of the collapsed cloud. We subsequently replace a
density-selected subset of SPH particles with stars by adopting a
local star-formation efficiency proportional to $\rho^{1/2}$.  As a
consequence, the local star formation efficiency exceeds 30 per cent,
whereas globally only a few per cent of the gas is converted to stars.
The stellar distribution by the time gas is converted to stars is very
clumpy, with typically a dozen bound conglomerates that consist of 100
to $10^4$ stars.  We continue to evolve the stars dynamically using
the collisional $N$-body method, which accurately treats all pairwise
interactions, stellar collisions and stellar evolution. We analyze the
results of the $N$-body simulations when the stars have an age of
2\,Myr and 10\,Myr.  During the dynamical simulations, massive
clusters grow via hierarchical merging of smaller clusters.  The shape
of the cluster mass function that originates from an individual
molecular cloud is consistent with a Schechter function with a
power-law slope of $\beta = -1.73$ at 2 Myr and $\beta = -1.67$ at 10
Myr, which fits to observed cluster mass function of the Carina
region.  The superposition of mass functions have a power-law slope of
$\aplt -2$, which fits the observed mass function of star clusters in
the Milky Way, M31 and M83. We further find that the mass of the most
massive cluster formed in a single molecular cloud with a mass of
$M_{\rm g}$ scales with $6.1 M_{\rm g}^{0.51}$ which also agrees with
recent observation of the GMC and young clusters in M51.

\end{abstract}


\section{Introduction}

Observed star forming regions show filamentary or spumous structure,
which appears to be a natural consequence of the star formation
process \citep{2010A&A...518L.102A}.  Star forming regions are thought
to be the results of the gravitational collapse of giant molecular
clouds \citep[][and references therein]{2007ARA&A..45..565M}.  Once
the stars have formed they start to develop a wind, and then the first
supernovae explosions occur a few million years later. These outflows
cause the residual gas to be blown away.  The stellar distribution as
a consequence will be super-virial after all the gas is lost, but
virial equilibrium is quickly re-established, after which the cluster
will have lost some stars and the remaining bound stars eventually
form a spherical and centrally concentrated distribution.  There are
numerous observed examples of such young clusters that are about to
emerge from their parental molecular cloud \citep[][and references
  therein]{2003ARA&A..41...57L}.

The initial collapse of the molecular cloud is also driven by
(magneto)hydrodynamical processes, radiation, and chemical reactions,
in contrast to the purely dynamical evolution of gas-deprived
clusters.  These processes have been studied extensively from an
observational point of
view~\citep{2006ApJ...637..850K,2007ARA&A..45..481Z,2008A&A...479L..25Z}
and
numerically~\citep{2008MNRAS.389.1556B,2010ApJ...711.1017P,2012MNRAS.419.3115B,2012ApJ...759....9K,2012ApJ...754...71K,2012ApJ...760..155K,2012ApJ...761..156F}.
During the transition phase from a gas dominated cluster to a purely
stellar cluster, the spectrum of physical processes broadens
dramatically and includes hydrodynamics, gravity, nuclear fusion and
radiation processes.  It is only at the later, gas deprived state,
that dynamical process start to dominate the evolution of the stellar
system.

Most numerical studies aim at one of these extremes, either the star
formation process
\citep[e.g.,][]{2003MNRAS.343..413B,2008MNRAS.389.1556B} or the
gas-deprived evolution of the star cluster starting from artificial
fractal initial conditions \citep[e.g.,][]{2014MNRAS.438..620P}.  In
some studies the final conditions of the former are used as initial
conditions for the latter
\citep{2010MNRAS.404..721M,2012MNRAS.425..450M}.  In
\citet{2010MNRAS.404..721M} and \cite{2012MNRAS.425..450M} the problem
is addressed in some discrete steps.  They start by performing a
hydro-dynamical simulation including sink-particles to follow the
growth in mass of two accreting point-masses which represent
proto-stellar objects.  After about one free-fall time of the
molecular cloud they instantaneously remove all the residual gas and
continue the simulation using a direct $N$-body code to study the
dynamical evolution of the gas-deprived stellar system.

Although removing the gas instantaneously is an extreme measure, their
approach appears to work well.  The densest parts of their stellar
distributions are relatively small 0.1--0.2 pc and already rather
deprived of residual gas
\citep{2012MNRAS.419..841K,2012MNRAS.425..450M}.  The relatively small
and dense stellar clumps therefore survive even under their rather
extreme assumption of instantaneous gas expulsion.

We aim to acquire a better understanding of the formation process of
star clusters, using a method similar to the one used in
\citet{2010MNRAS.404..721M}. We decide to adopt a lower resolution to
avoid using sink particles.  As in \citet{2010MNRAS.404..721M}, we
start by performing a hydrodynamical simulation for about a free-fall
time, after which we analyze the results and generate stars. The
residual gas is removed instantaneously, and we continue the
simulation by means of direct $N$-body integration.  Instead of using
sink-particles, we convolve gas particles to stars adopting a simple
star-formation efficiency (SFE) that depends on the local free-fall
time.  This procedure is considerably easier to implement and requires
much less computer time, and therefore we can treat clusters with a
mass of $>10^4M_{\odot}$, that was not possible in previous
simulations.  Such a method also allows us to have more control over
the star formation process, and it makes the interpretation of the
result more transparent.

In our experiments, we are particularly interested in the earliest
dynamical evolution of the clusters, as it formed in a collapsed
molecular cloud after the gas expulsion. We study this dynamical
evolution up to about 2\,Myr for clusters with a mass of $\sim 10^4
\,M_{\odot}$. We refer to these clusters as young massive
clusters. Our study is motivated by the recent finding of the
accretion of a smaller stellar clump in R136
\citep{2012ApJ...754L..37S}, but we are also intrigued by the
formation processes of Westerlund 1, Westerlund 2 and NGC 3603,
because recent studies using $N$-body simulations suggest that they
formed via mergers of smaller sub-clusters
\citep{2012ApJ...753...85F}.

\section{Numerical methods}
In our simulations we combine a smoothed particles hydrodynamics 
(SPH) code with a direct $N$-body code.  
The method consists of three steps: 
\begin{enumerate}
\item
Perform an SPH simulation of a turbulent molecular cloud for an
initial free-fall time scale. 
\item
Convert gas particles to stellar particles assuming a star formation
efficiency (SFE), which depends on the local gas density, and remove the
residual gas (SPH particles).
\item
Turn on the direct $N$-body simulation for integrating the equations
of motion of the stellar particles.
\end{enumerate}
The initial conditions for SPH simulations are taken from an
isothermal homogeneous gas sphere with a turbulent velocity field 
\citep{2003MNRAS.343..413B} with a spectral index of $k=-3$. 
The size and
total mass for our standard model are 10 pc and $4\times
10^5M_{\odot}$, respectively. The mean gas density is then 
$\sim 100M_{\odot}{\rm pc}^{-3}$($\sim 1700\,{\rm cm}^{-3}$ assuming that 
the mean weight per particle is $2.33m_{\rm H}$) and the free-fall time 
$t_{\rm ff} = 0.83$ Myr.

\subsection{The hydrodynamical collapse of the molecular cloud}

We initialize the gas cloud by giving it zero total energy (potential
plus kinetic). We perform additional simulations with a mean gas
density of $\sim 10M_{\odot}{\rm pc}^{-3}$ ($\sim 170\,{\rm cm}^{-3}$)
by increasing the overall dimension of the system and adopting a total
mass of $4\times 10^4$--$2\times 10^6M_{\odot}$.  With a mean gas
density of $\sim 10M_{\odot}{\rm pc}^{-3}$, $t_{\rm ff} =
2.5$\,Myr. We adopt a gas temperature of 30K.  We summarize the
initial conditions of the simulations in Table \ref{tb:models_hydro}.
For each initial condition, we perform 1--3 runs with different random
seeds for the generation of the turbulence in order to see the
run-to-run variation. The number of random seeds are indicated as 's'
in the table.

For our SPH simulations, we adopt $1M_{\odot}$ per particle.  The SPH
softening length ($h$) is chosen such that $\rho h^3 = m N_{\rm nb}$
\citep{2002MNRAS.333..649S}, where $\rho$ is the density, $m$ and the
particle mass, and $N$ is the target number of neighbor particles. We
adopt $N_{\rm nb}=64$, with which the mass resolution is $\sim
100M_{\odot}$. With this setup, the minimum scale we can resolve
the Jeans instability is $h = 1.7$\,pc. These mass and size are 
comparable to
the typical ($\sim 1$ pc) size of embedded clusters
\citep{2003ARA&A..41...57L}.  Our SPH simulations therefore cannot
resolve the formation of individual stars.  The gravitational
softening length is taken to be constant (approximately the smallest
softening length encountered in the simulation) $\epsilon_{\rm grav} =
0.1$ pc for the gas particles in order to improve energy and momentum
conservation.  The resolution of our simulations are lower than recent
simulations which aim at simulating star formation in turbulent
molecular clouds \citep{2011MNRAS.410.2339B,2012MNRAS.419.3115B}. We
motivate this relatively low resolution by our aim at reproducing the
clumpy structure of star forming regions, rather than the microscopic
details of the star formation process.  Of course, ideally we would
like to resolve also the latter, but this is currently unpractical, at
least, and we think that our approach is reasonable and a considerable
advance over adopting a simple virialized Plummer sphere
\citep{1911MNRAS..71..460P} or King model \citep{1966AJ.....71...64K}
to generate the initial realization for the gravitational $N$-body
simulations.

The generation of the initial conditions and the SPH simulations are
performed with the Astronomical Multipurpose Software Environment
(AMUSE) \citep{2013CoPhC.183..456P,2013A&A...557A..84P}\footnote{see
  \url{http://amusecode.org/}.}.  AMUSE is a follow-up of the earlier
MUSE environment \cite{2009NewA...14..369P}, which was intended to be
a general purpose framework for performing large scale astronomical
simulations.  In the AMUSE framework we adopted the {\tt Python}
programming language to encapsulate the fundamental physics solvers,
which are written in high-performance compiled computer code. The
flexibility of AMUSE makes it possible to perform simulations with one
implementation of the numerical solver, and then repeat the same
calculation with a different solver by changing one line in the {\tt
  AMUSE-python} script.

As the engine for performing the hydrodynamical simulation we adopted
the SPH code {\tt Fi}
\citep{1989ApJS...70..419H,1997A&A...325..972G,2004A&A...422...55P,
  2005PhDT........17P}. Within {\tt Fi} the internal time step is
controlled internally, but we constrain this timestep from within the
framework to 0.025 Myr.

\begin{figure}
\begin{center}
\includegraphics[width=\columnwidth]{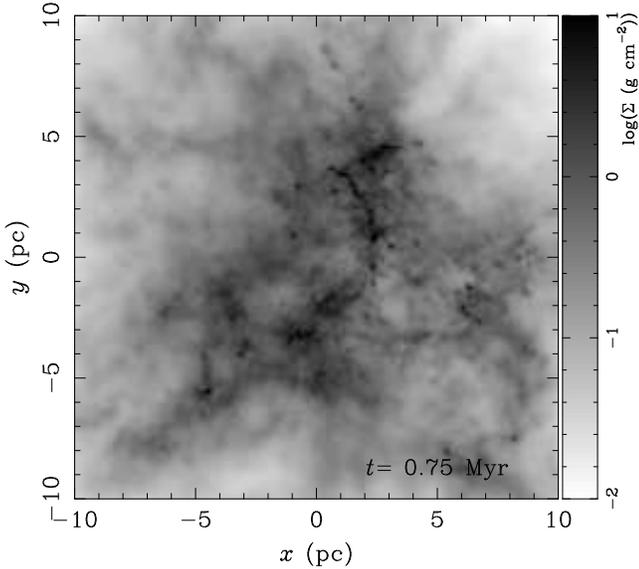}
\end{center}
\caption{Gas surface density at an age of $0.9t_{\rm ff}$ for model
 m400k-d100-30K. \label{fig:gas_star}}
\end{figure}

We stop the hydrodynamical simulation at $0.9t_{\rm ff}$.  By this
time the volume density reaches $10^{7-8} {\rm cm}^{-3}$
($10^{6-7}M_{\odot} \mathrm{pc}^{-3}$) and the surface density reaches
a value of $10 \mathrm{g}\,\mathrm{cm}^{-2}$ 
($10^5M_{\odot} {\rm pc}^{-2}$).  In Figure\,\ref{fig:gas_star} we present 
an example of the gas density
distribution at the moment we stop the hydrodynamical calculations.

\begin{table*}
\begin{center}
\caption{Models\label{tb:models_hydro} for hydrodynamical simulations ('s' indicates the random seeds for the turbulence).}
\begin{tabular}{lccccccc}\hline \hline
Model  & Total mass & $N$ of particles & Radius  & Density & Thermal Energy & Temperature\\
      & $M_{\rm g}(M_{\odot})$ & $N_{\rm g}$ &$r_{\rm g}$(pc)  & $\rho_{\rm g}({\rm cm}^{-3})$ & $E_{\rm t}/E_{\rm k}$ & $T$ (K) \\ \hline
m2M-d10-s18-30K  & $2\times 10^6$ & $2\times 10^6$& 35.9  & 170  & $1.1\times10^{-3}$ & 30 \\

m1M-d100-s7-30K  & $1\times 10^6$ & $1\times 10^6$& 13.4  & $1.7\times10^3$  & $8.4\times10^{-4}$ & 30 \\
m1M-d100-s12-30K & $1\times 10^6$ & $1\times 10^6$& 13.4  & $1.7\times10^3$  & $8.4\times10^{-4}$ & 30 \\
m1M-d100-s10-30K & $1\times 10^6$ & $1\times 10^6$& 13.4  & $1.7\times10^3$  & $8.4\times10^{-4}$ & 30 \\

m1M-d10-s3-30K & $1\times 10^6$ & $1\times 10^6$& 28.5  & 170  & $1.7\times 10^{-3}$ & 30   \\
m1M-d10-s4-30K & $1\times 10^6$ & $1\times 10^6$& 28.5  & 170  & $1.7\times 10^{-3}$ & 30   \\

m400k-d100-s1-30K  & $4\times 10^5$ & $4\times 10^5$& 10.0 & $1.7\times10^3$  & $1.6\times10^{-3}$ & 30  \\
m400k-d100-s5-30K  & $4\times 10^5$ &$4\times 10^5$& 10.0  & $1.7\times10^3$  & $1.6\times10^{-3}$ & 30 \\
m400k-d100-s11-30K & $4\times 10^5$ &$4\times 10^5$& 10.0  & $1.7\times10^3$  & $1.6\times10^{-3}$ & 30 \\

m400k-d10-s8-30K   & $4\times 10^5$ &$4\times 10^5$ & 21.0 & 170  & $3.3\times10^{-3}$ & 30  \\
m400k-d10-s9-30K   & $4\times 10^5$ &$4\times 10^5$ & 21.0 & 170  & $3.3\times10^{-3}$ & 30  \\

m100k-d100-s2-30K  & $1\times 10^5$ & $1\times 10^5$ & 6.2 & $1.7\times10^3$  & $3.9\times10^{-3}$ & 30\\
m100k-d100-s6-30K  & $1\times 10^5$ & $1\times 10^5$ & 6.2 & $1.7\times10^3$  & $3.9\times10^{-3}$ & 30\\
m100k-d100-s13-30K & $1\times 10^5$ & $1\times 10^5$ & 6.2 & $1.7\times10^3$  & $3.9\times10^{-3}$ & 30\\

m40k-d100-s20-30K   & $4.1\times 10^4$ & $4.1\times 10^4$& 4.6 & $1.7\times10^3$  & 0.0071 & 30  \\
m40k-d100-s21-30K   & $4.1\times 10^4$ & $4.1\times 10^4$& 4.6 & $1.7\times10^3$  & 0.0071 & 30  \\
m40k-d100-s22-30K   & $4.1\times 10^4$ & $4.1\times 10^4$& 4.6 & $1.7\times10^3$  & 0.0071 & 30  \\
\hline
\end{tabular}
\medskip
\end{center}
\end{table*}

\subsection{Forming the stars}

\begin{figure}
\begin{center}
\includegraphics[width=70mm]{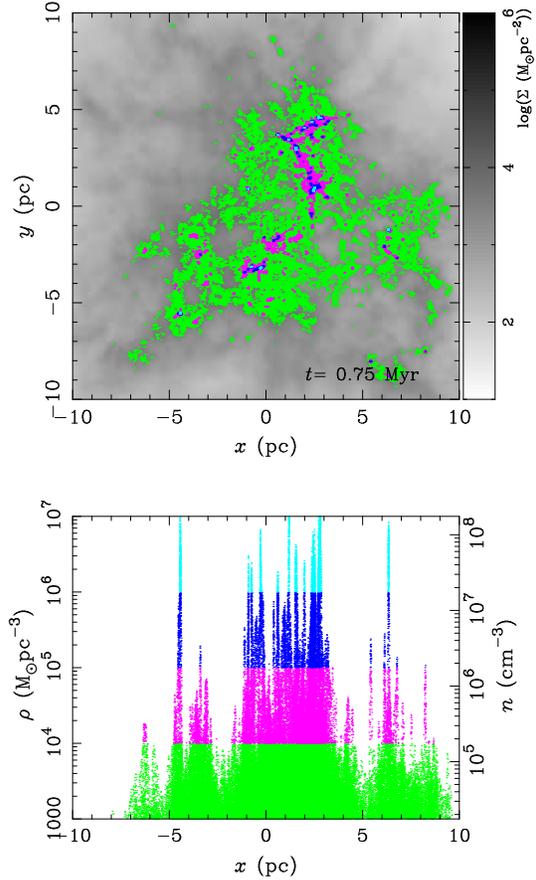}
\end{center}
\caption{Gas surface density at $0.9t_{\rm ff}$ for model m400k-d100-30K
  (top panel in gray scale). Colors indicate dense region; cyan, blue,
  magenta, green indicates regions with the volume density of $>10^6$,
  $10^{5-6}$, $10^{4-5}$, and $10^{3-4} M_{\odot}$pc$^{-3}$,
  respectively. Gas volume density distribution projected on $x$-axis
  (bottom).  Colors are the same as in top panel.\label{fig:gas_sd}}
\end{figure}

We continue by analyzing the resulting density distribution of the
collapsed molecular gas cloud.  In Figure \ref{fig:gas_sd} we present
a projected image of the density distribution of the gas at an age of
0.75\,Myr ($\sim 0.9t_{\rm ff}$) after the start of the hydrodynamical
simulation.  The densest regions reached $\sim 10^{6} M_{\odot}{\rm
  pc}^{-3}$, which is consistent to the results from earlier SPH
simulations that included star formations through sink particles
\citep{2010MNRAS.404..721M}.

The conversion of the SPH particles to stars was realized by adopting
a local star formation efficiency, $\epsilon_{\rm loc}$, which we
calculate using
\begin{eqnarray}
  \epsilon_{\rm loc} &=& \alpha_{\rm sfe} 
                       \sqrt{\frac{\rho}{10^2 (M_{\odot}{\rm pc}^{-3})}}\\
                   &=& \alpha_{\rm sfe} 
                       \sqrt{\frac{\rho}{1.7\times 10^3 ({\rm cm}^{-3})}}.
\label{eq:eff}
\end{eqnarray} 
Here $\rho$ is the local volume density, which is measured at the
location of each individual SPH particle in the simulation.  The
coefficient, $\alpha_{\rm sfe}$, controls the star formation
efficiency and is a free parameter in our simulations. With this
prescription, the local SFE correlates with the instantaneous
free-fall time of the gas via the square-root of the gas density.
This assumption is motivated by recent result that indicate that the
star formation rate scales with the free-fall time
\citep{2012ApJ...745...69K}.  We adopted $\alpha_{\rm sfe}=0.02$ for
the models with $\rho_{\rm g}=100M_{\odot}{\rm pc}^{-3}$, which is
similar to what was obtained by \citet{2012ApJ...745...69K}.  We chose
a higher value of $\alpha_{\rm sfe}= 0.04$ for $\rho_{\rm g}=170{\rm
  cm}^{-3}$, because in these cases the time for a part of the system
to evolve from the moment when it reaches $1700{\rm cm}^{-3}$
to the end of the hydrodynamical simulation is twice as long as the
free-fall time of the models with $1700{\rm cm}^{-3}$.

\begin{figure}
\begin{center}
\includegraphics[width=80mm]{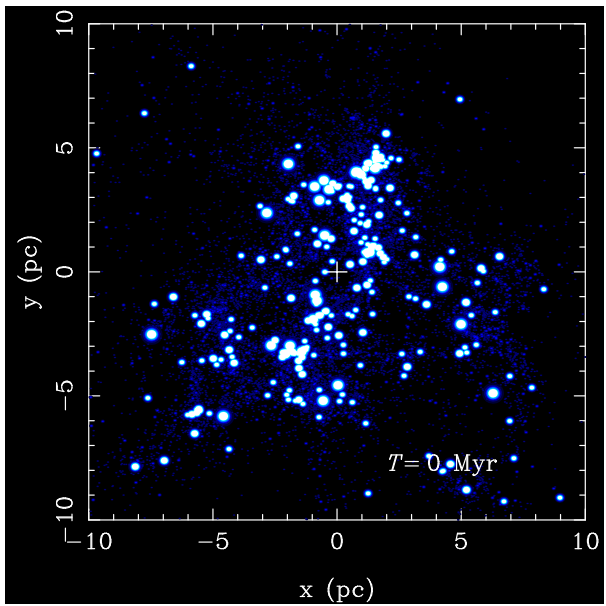}
\end{center}
\caption{The initial spatial distribution of the stars 
  for the $N$-body simulation.\label{fig:stars}}
\end{figure}

We replace the densest SPH particles with stellar particles by
adopting the local star formation efficiency of equation
(\ref{eq:eff}).  This resulted in the star formation efficiency in the
dense regions ($\rho \geq 1.7\times 10^4 {\rm cm}^{-3} $ i.e.,
$10^{3}M_{\odot} {\rm pc}^{-3}$), $\epsilon_{\rm d}$, of 20--30\%
which is consistent with the observed star formation efficiency
\citep{2003ARA&A..41...57L}. The residual gas is assumed to be ejected
from the system. In Figure\,\ref{fig:stars} we present the stellar
distribution that is obtained from the hydrodynamical simulation after
the SPH particles have been converted to stars and the residual gas is
removed.

The positions and velocities of the stars are identical to those of
the SPH particles from which they formed.  The mass of a star was
selected randomly from the \cite{1955ApJ...121..161S} mass function
between a minimum and maximum mass of 0.3\,$M_\odot$ and 100
$M_{\odot}$. The mean mass of this distribution turns out to be
$1M_{\odot}$.  As a consequence, this method does not conserve mass
locally, but globally mass is conserved. After removing all SPH
particles, the residual stellar system was super-virial with a virial
ratio $Q_{\rm vir}>1$. We did not generate any primordial binaries in
the initial conditions because during our integration time ($\aplt
10$Myr) hard binaries which are tight enough to affect on the dynamical
evolution of the host clusters interact only very rarely with the
surrounding stars due to their small separations
\citep{2011Sci...334.1380F}.

The global star formation efficiency, $\epsilon$, for the entire
molecular structure turned out to be a few per cent. This fraction is
a result of the simulations and the adopted local star formation
efficiency.  In Table \ref{tb:models_nbody} we give the values for
$\epsilon$ and $\epsilon_{\rm d}$ obtained from our simulations.

\begin{table*}
\begin{center}
\caption{Models for $N$-body simulations \label{tb:models_nbody}}
\begin{tabular}{lccccc}\hline \hline
Model  & Total mass & $N$ of particles  & Virial ratio & SFE (Global) & SFE (Dense)\\
     & $M_{\rm s}(M_{\odot})$ & $N_{\rm s}$  & $Q_{\rm vir}$  & $\epsilon$ & $\epsilon_{\rm d}$\\ \hline
m2M-d10-s18-30K & $1.0\times 10^5$ & 99546 &  3.9  & 0.050 & 0.39 \\

m1M-d100-s7-30K  & $1.1\times 10^5$ & 109952 & 6.1  & 0.11 & 0.26\\

m1M-d100-s12-30K  & $9.4\times 10^4$ & 94464 & 2.2  & 0.096 & 0.23\\

m1M-d10-s3-30K & $7.8\times 10^4$ & 78201 &  0.66  & 0.079 & 0.54 \\
m1M-d10-s4-30K & $5.7\times 10^4$ & 56681 &   2.2   & 0.058 &  0.41\\

m400k-d100-s1-30K  & $3.0\times 10^4$ & 30487 & 2.4  & 0.074 & 0.21 \\
m400k-d100-s5-30K  & $2.4\times 10^4$ & 24211 & 4.8  & 0.059 & 0.16 \\
m400k-d100-s11-30K & $4.0\times 10^4$ & 39933 & 1.5 & 0.074 & 0.26 \\

m400k-d10-s8-30K & $1.9\times 10^4$ & 19236 &  2.8 & 0.047 & 0.39 \\
m400k-d10-s9-30K & $2.0\times 10^4$ & 19672 &  2.6 & 0.047 & 0.36 \\

m100k-d100-s2-30K  & $1.3\times 10^4$ & 12833 &  0.58  & 0.13 & 0.34 \\
m100k-d100-s6-30K  & $4.6\times 10^3$ & 4572  &  7.8  & 0.045 & 0.14 \\
m100k-d100-s13-30K & $8.0\times 10^3$ & 7987  &  1.9  & 0.079 &  0.22  \\

m40k-d100-s20-30K   & $2.9\times 10^3$ & 2866  & 1.7 & 0.079 & 0.21\\
m40k-d100-s21-30K   & $3.2\times 10^3$ & 3175  & 1.9 & 0.077 & 0.21\\
m40k-d100-s22-30K   & $2.1\times 10^3$ & 2073  & 2.0 & 0.051 & 0.16\\

\hline
\end{tabular}
\medskip

\end{center}
\end{table*}

\subsection{The dynamical evolution of the cluster}

We now use the stellar masses, positions, and velocities as initial
realizations for our $N$-body calculations in which we study the
dynamical evolution of the stellar system.  For convenience we
associate the moment at which we start the $N$-body simulations with
$t=0$\,Myr.  The forces between each pair of stars is calculated
directly, and the numerical integration of the equations of motion was
performed using the sixth-order Hermite scheme
\citep{2008NewA...13..498N}.  The $N$-body code runs without softening
and with a time step parameters $\eta = 0.1$--0.3. The energy error was 
less than 0.1\% for all simulations. It is small enough for obtaining a 
scientifically interpretable result in such $N$-body simulations
\citep{2014ApJ...785L...3P}.

If two stars approach each other closer than the sum of their radii,
we resolve the collision by summing the mass and conserving the
angular momentum.  The stellar radius was calculated using the
description in \citep{2000MNRAS.315..543H,2012A&A...546A..70T} and for
stars $>100M_{\odot}$ we extrapolated the results \citep[see][for the
  details]{2009ApJ...695.1421F,2012ApJ...753...85F}.  We only adopted
very simple stellar evolution prescription in which a star turns into
a black hole directly after the main sequence
\citep{2000MNRAS.315..543H}.  The supernovae were assumed to be
symmetric and therefore no natal kick was delivered to the compact 
remnant. By the end of the simulations (10 Myr), stars with a mass of 
$\apgt 20M_{\odot}$ reaches the end of their main-sequence life-time and
evolve to black holes.

\begin{table*}
\begin{center}
\caption{The results of simulations at $t=2$Myr\label{tb:results}.
  $M_{\rm s, cl}/M_{\rm s}$ is the stellar mass fraction which belongs
  to clusters. $M_{\rm c, max}$ is the mass of the most massive
  clusters formed in the simulations. $N_{\rm c}$ is the number of
  clusters.  $\beta_1$ is the power of the fitted cluster mass
  function adopting the value of $M_{\rm c, max}$ obtained from the
  simulations. $\beta_2$ and $A$ are the power and the factor of the
  fitted cluster mass function (equation (\ref{eq:fit})) but with
  $M_{\rm c, max}=0.20M_{\rm g}^{0.76}$. The fitting results exist
  only for models with $N_{\rm c}>4$.  Averaging the results, we
  obtain $\beta_1=-1.71\pm 0.18$, $\beta_2 = -1.73\pm 0.17$, and
  $A=0.64\pm 0.29$.  }
\begin{tabular}{lcccccc}\hline \hline

Model & $ M_{\rm s, cl}/M_{\rm s}$ & $M_{\rm c, max}(M_{\odot})$ & $N_{\rm c}$ & $\beta_1$ & $\beta_2$ & $A$   \\ \hline
m2M-d10-s18-30K   & 0.31 & $3.1\times 10^4$ & 61 & $-1.68\pm 0.02$ & $-1.60\pm 0.08$ & $0.93\pm 0.15$ \\

m1M-d100-s7-30K  & 0.43 &  $9.5\times 10^3$ & 51 & $-1.65\pm 0.02$ & $-1.64\pm 0.02$ & $1.13\pm 0.10$\\ 
m1M-d100-s12-30K & 0.40 &  $8.9\times 10^3$ & 50 & $-1.74\pm 0.02$ & $-1.73\pm 0.02$ & $0.72\pm 0.05$\\  

m1M-d10-s3-30K   & 0.42 & $2.4\times 10^4$ & 30 & $-2.16\pm 0.04$ & $-2.14\pm 0.04$ & $0.09\pm 0.01$ \\
m1M-d10-s4-30K   & 0.33 & $2.5\times 10^3$ & 45 & $-1.93\pm 0.04$ & $-2.00\pm 0.03$ & $0.26\pm 0.03$ \\

m400k-d100-s1-30K & 0.33 & $3.0\times 10^3$ & 15 & $-1.53\pm 0.05$ & $-1.55\pm 0.05$ & $0.73\pm 0.12$ \\ 
m400k-d100-s5-30K & 0.40 & $2.7\times 10^3$ & 16 & $-1.59\pm 0.07$ & $-1.62\pm 0.07$ & $0.82\pm 0.16$ \\\ 
m400k-d100-s11-30K & 0.37 & $7.4\times 10^3$ & 14 & $-1.64\pm 0.04$ & $-1.58\pm 0.04$& $0.78\pm 0.11$  \\

m400k-d10-s8-30K & 0.38 & $1.4\times 10^3$ & 18 & $-1.63\pm 0.05$ & $-1.71\pm 0.04$& $0.50\pm 0.06$\\ 
m400k-d10-s9-30K & 0.36 & $2.1\times 10^3$ & 17 & $-1.71\pm 0.06$ & $-1.75\pm 0.06$ & $0.46\pm 0.08$\\ 

m100k-d100-s2-30K & 0.41 & $5.0\times 10^3$ & 4 & - & - & -\\ 
m100k-d100-s6-30K & 0.35 & $3.9\times 10^2$ & 7 & $-1.46\pm 0.20$ & $-1.81\pm 0.22$ & $0.38\pm 0.17$\\ 
m100k-d100-s13-30K & 0.45 & $8.5\times 10^2$ & 11 & $-1.51\pm 0.08$ & $-1.60\pm 0.08$ & $0.08\pm 0.16$\\ 

m40k-d100-s20-30K & 0.35 & $4.9\times 10^2$ & 4 & - & - & - \\ 
m40k-d100-s21-30K & 0.28 & $3.2\times 10^2$ & 2 & - & - & - \\ 
m40k-d100-s22-30K & 0.28 & $8.5\times 10^2$ & 2 & - & - & - \\ 

\hline
\end{tabular}
\end{center}
\medskip
\end{table*}

\begin{table*}
\begin{center}
\caption{The results of simulations at 10Myr\label{tb:results_10Myr}.
  Averaging the results, we obtain $\beta_1=-1.53\pm 0.16$, $\beta_2 =
  -1.67\pm 0.30$, and $A=0.63\pm 0.30$.  Here, we assume $M_{\rm c,
    max}=0.20M_{\rm g}^{0.76}$.  }
\begin{tabular}{lccccc}\hline \hline
Model    & $M_{\rm c, max}(M_{\odot})$   & $N_{\rm c}$& $\beta_1$ & $\beta_2$ & $A$ \\ \hline
m400k-d100-s1-30K & $5.3\times 10^3$ & 9 & $-1.41\pm 0.05$ & $-1.35\pm 0.06$ & $0.84\pm 0.16$\\ 
m400k-d100-s5-30K & $2.3\times 10^3$ & 14 & $-1.60\pm 0.05$ & $-1.58\pm 0.04$ & $0.83\pm 0.10$\\ 
m100k-d100-s2-30K & $5.2\times 10^3$ & 4 & -               & - & -\\ 
m100k-d100-s6-30K & $4.6\times 10^2$ & 5 &  $-1.75\pm 0.08$ & $-2.07\pm 0.08$ & $0.20\pm 0.03$\\ 

\hline
\end{tabular}
\end{center}
\medskip
\end{table*}

\section{Cluster Mass Function }

\subsection{Cluster Finding}

At an age of 2\,Myr and at 10\,Myr, we interrupt the simulations to
study the clumps in the stellar distribution.  We identify these
clumps as star clusters.

These clusters are detected using the HOP \citep{1998ApJ...498..137E}
clump finding algorithm (which is also incorporated in the AMUSE
framework).  The outer cut-off density (somewhat related to the
density of an individual clump) was set to 
$\rho_{\rm out} = 4.5M_{\rm s}/(4\pi r_{\rm h}^{3})$, which is three times the half-mass density
of the entire system.  Other parameters in HOP are the number of
particles to calculate the local density, for which we adopted $N_{\rm
  dense}=64$, the number of particles for neighbor search ($N_{\rm
  hop}=64$) and the number of particles of neighbors to determine for
two groups to merge ($N_{\rm merge}=4$).  Because star clusters have a
relatively high density contrasts compared to dark matter halos, to
which the method is applied with the default parameters, we adopted
$8\rho_{\rm out}$ for the saddle density threshold and $10\rho_{\rm
  out}$ for the peak density threshold. We do not include clusters
with fewer than 64 stars in our analysis, because both the SPH
simulations and the clump finding method used for analysis cannot
resolve them.  Such small clusters hardly ever exceed $100M_{\odot}$,
which is consistend with the generally adopted minimum mass for a star
cluster, regardless the arbitraryness of this choise. We adopted these
parameter because HOP was most successful in detecting all the
clusters, but verified that changing the parameters does not influence
our results qualitatively.  The clumps detected are not necessarily
bound, but we detected the members based on geometry. This might not
be what a theorist normally would identify as a cluster, but from an
observational point of view it is often hard to separate out the
unbound stars from the bound stars. With the adopted method we mimic
an observational identification criterion.

\begin{figure*}
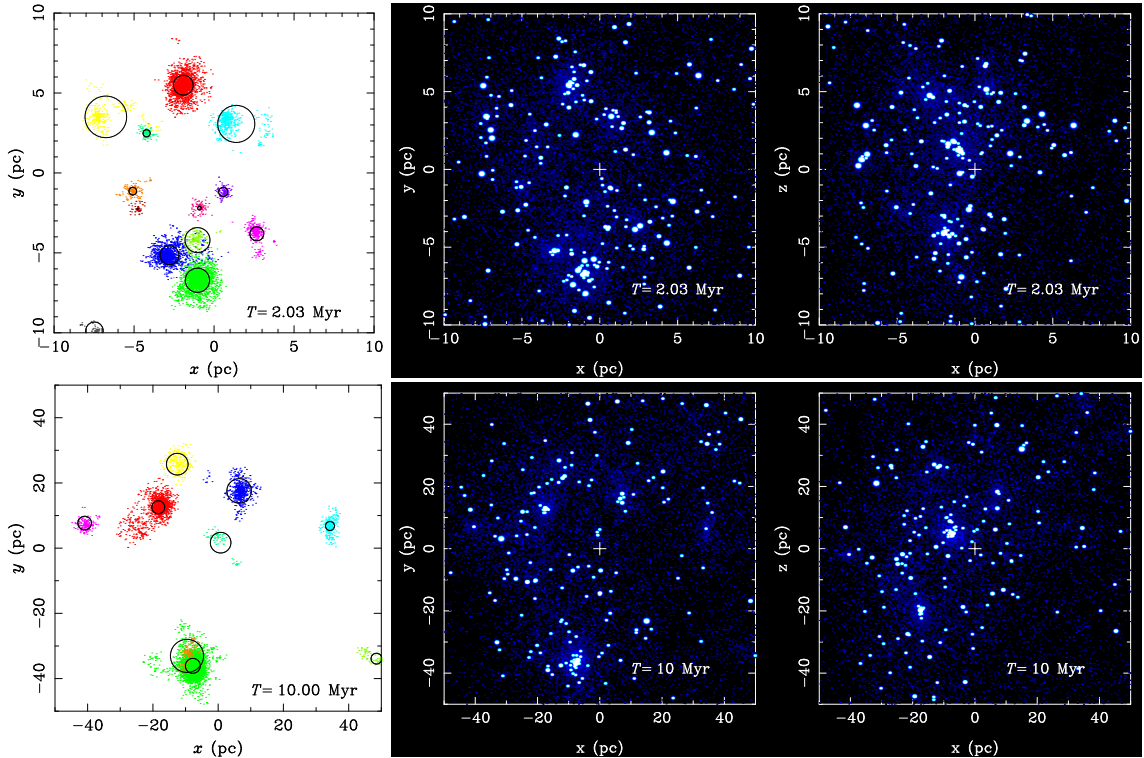

\begin{center}
\includegraphics[width=50mm]{f4a.eps}
\includegraphics[width=100mm]{f4b.eps}\\
\includegraphics[width=50mm]{f4c.eps}
\includegraphics[width=100mm]{f4d.eps}
\end{center}
\caption{Projected snapshots of model m400k-d100-30K at 2\,Myr (top)
  and at 10\,Myr (bottom).  The results of the clump finding algorithm
  are presented in the left-most two panels. Each clump is presented
  in a single color, and the black circles show the respective
  half-mass radii of the clumps. Overlaying circles are separated in
  the third dimension. To the right we present two projections
  ($x$-$y$ in the middle and $y$-$z$ in the right) of all the stars in
  the simulation.
 \label{fig:clump_finding}}
\end{figure*}

After identifying all clusters, we determine their total mass and
half-mass radius. In Figure \ref{fig:clump_finding} we present two
examples at 2\,Myr and at 10\,Myr of the clusters identified using
this procedure.

\subsection{The Star-Cluster Mass Function}

\begin{figure*}
\begin{center}
\includegraphics[width=120mm]{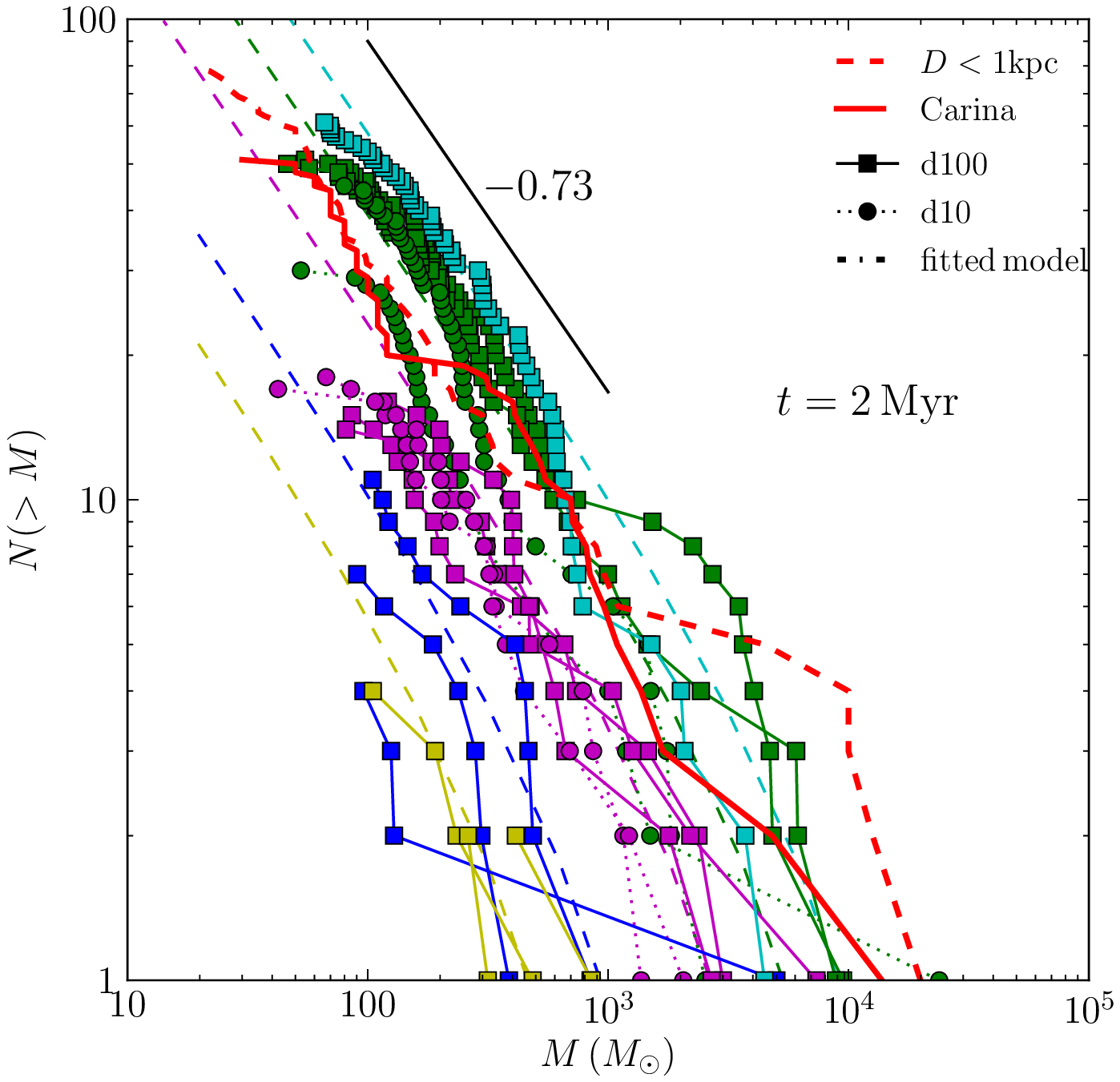}
\end{center}
\caption{Cumulative cluster mass function obtained from the
  simulations with a temperature of 30K when the stars had an age of 2
  Myr.  The colors represent different masses of initial molecular
  clouds: cyan, green, magenta, blue, and yellow indicate $M_{\rm g}$
  of $2\times 10^6, 10^6, 4\times 10^5, 10^5$ and $4\times 10^4
  M_{\odot}$, respectively.  The dashed curves give the fitted MF (see
  equation (\ref{eq:fit})).  We adopt $A=0.64$, $\beta=-1.73$, and
  $M_{\rm c, max}=0.20M_{\rm g}^{0.76}$.  The red dashed curve is the
  mass function of the MW young clusters ($<3$ Myr) within 1 kpc from
  the Sun (indicated by $D<1$kpc). The data is from
  \citep{2003ARA&A..41...57L,
    2008A&A...487..557P,2010ARA&A..48..431P}.  We here assumed that
  all the (embedded) clusters in \citet{2003ARA&A..41...57L} are
  younger than 3 Myr. Red thick curve is the cluster mass function of
  the Carina region \citep{2011ApJS..194....9F}.
  \label{fig:CMF_sim}}
\end{figure*}

For each simulation we can now construct a mass function (MF) of 
star clusters.  In Figure \ref{fig:CMF_sim} we present the cumulative
mass distribution for star clusters in our simulations.  The offset
scatter among the models is large, but the power of the MF is 
similar, irrespective of the model.  We compare the obtained MFs with
the Schechter function:
\begin{eqnarray}
  \phi (M) \equiv \frac{dN}{dM} \propto M^{\beta}\exp\left(
                 -\frac{M}{M_{\rm cut}} \right).
\label{Eq:Schechter}
\end{eqnarray}
Here $M$ is the mass of clusters and $M_{\rm cut}$ is the cut-off
mass. Integration of equation (\ref{Eq:Schechter}) results in the
cumulative MF, which has the form
\begin{eqnarray}
  N(>M)\propto M^{\beta+1}\exp\left( -\frac{M}{M_{\rm cut}} \right).
\label{eq:cum}
\end{eqnarray}
We use equation (\ref{eq:cum}) to fit (using least mean squares) the
mass distribution of the clusters we obtained in each of our
simulations.  The average of the best fit parameter for all models
together is $\beta = -1.55 \pm 0.41$ for the slope of the mass
function.  The uncertainty in the fitting procedure for detarmining
  $M_{\rm cut}$ is very large.

Even though we were unable to determine a reliable value for $M_{\rm cut}$, 
we can use the mass of the most massive cluster ($M_{\rm c, max}$) as 
$M_{\rm cut}$. 
Scaling equation (\ref{eq:cum}) by assuming that $M_{\rm cut}=M_{\rm c, max}$
, the cumulative mass function becomes
\begin{eqnarray}
  N(>M)= \frac {A M^{\beta+1}\exp\left( -\frac{M}{M_{\rm c, max}} \right)} 
               {M_{\rm c, max}^{\beta+1}\exp\left( -1 \right)}. 
\label{eq:fit}
\end{eqnarray}
Here $A$ is a factor and we scaled in such a way that $N=A$ for
$M=M_{\rm c, max}$.  Fitting this function to the cluster mass
functions results in $\beta = -1.71\pm 0.18$.  In
Table~\ref{tb:results} we give for each model the maximum cluster mass
and best fit parameters.

\begin{figure}
\begin{center}
\includegraphics[width=80mm]{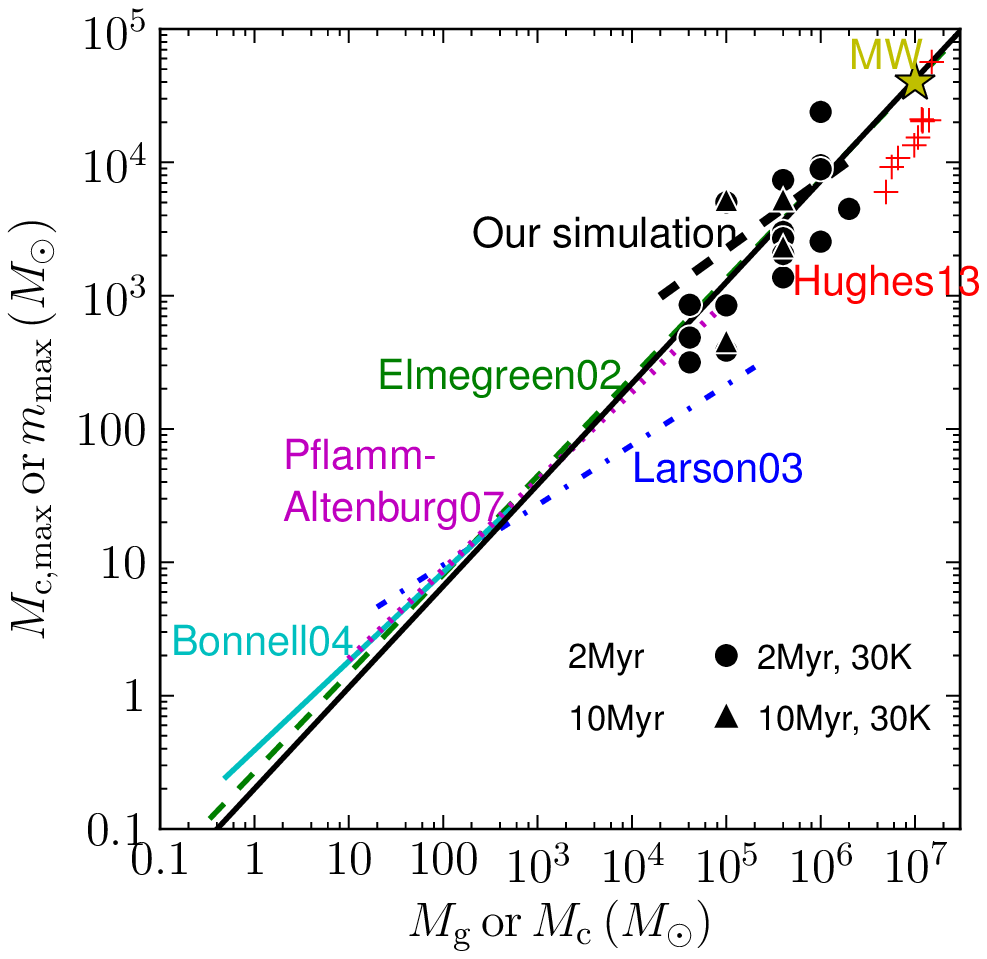}
\end{center}
\caption{The relation between the mass of the most massive cluster
  ($M_{\rm c, max}$) and the total mass of the molecular cloud
  ($M_{\rm g}$) or between the mass of the most massive star ($m_{\rm
    max}$) and the total mass of its host cluster ($M_{\rm c}$). Black
  dots and triangles indicate the results of our simulations with 30K
  for $t=$2\,Myr and 10\,Myr, respectively. Thick dashed line
  indicates the fitted function to our result ($M_{\rm c,
    max}=6.3M_{\rm g}^{0.51}$), and thick full line is also fitted but
  forced to at $m_{\rm s}=0.01 M_{\odot}$ at $M_{\rm g}=0.02
  M_{\odot}$ ($M_{\rm c, max}=0.20M_{\rm g}^{0.76}$).  Red pluses
  indicate the observed relation between the mass of the most massive
  GMCs and clusters in different environments in M51
  \citep{2013ApJ...779...44H}. Star indicates the most massive cluster
  and GMC in the MW.  The color thin dashed, dash-dotted, full, and
  dotted lines give the relation between the mass of the most massive
  star as a function of its host cluster mass; $m_{\rm s, max}=0.27
  M_{\rm c}^{0.74}$ \citep{2002ApJ...577..206E}, $m_{\rm s, max}=1.2
  M_{\rm c}^{0.45}$ \citep{2003ASPC..287...65L}, $m_{\rm s, max}=0.3
  M_{\rm c}^{2/3}$ \citep{2004MNRAS.349..735B}, and $m_{\rm s,
    max}=0.4 M_{\rm c}^{0.67}$\citep{2007ApJ...671.1550P},
  respectively (see also \citet{2010MNRAS.401..275W}). The length of
  each line is proportional to the mass range from which the relation
  is obtained.
  \label{fig:m_max}}
\end{figure}

There appears to be a clear relation between the most massive cluster
($M_{\rm c, max}$) and the initial total mass in gas ($M_{\rm g}$),
which we fitted using least squares to the form $M_{\rm c,
  max}=6.3M_{\rm g}^{0.51}$.  In Figure~\ref{fig:m_max} we over plot
the results of the simulations with the fitted function (black thick
dashed line).  To compare with observations, we over-plot recent
results of the mass of the most massive GMCs and star clusters in
different regions in M51 \citep{2013ApJ...779...44H}. If we consider
that the most massive GMC forms the most massive star cluster, this
observational result is directly comparable to our results.
Interestingly, this relation is also quite consistent with the
relation between the mass of the most massive star ($m_{\rm max}$) and
its host cluster ($M_{\rm c}$) found in observations
\citep{2003ASPC..287...65L,2007ApJ...671.1550P}, simulations
\citep{2004MNRAS.349..735B}, and theoretical models
\citep{2002ApJ...577..206E}. We make the connection between this
relation and stellar scales by adopting a minimum stellar mass (brown
dwarf) of $0.01M_{\odot}$ which with a 50\% SFE must have formed from
a gas cloud with a mass of $0.02M_{\odot}$. When we include this point
as a boundary condition in the fitting procedure, we obtain $M_{\rm
  c,max}=0.20M_{\rm g}^{0.76}$ as a
best fit to the data (including this artificial point). This result
does not strongly depend on the value of the artificial point. If we
adopt the stellar mass of $0.5M_{\odot}$ and the parental gas mass of
$2.15M_{\odot}$, which is obtained from the relation between the gas
mass and the stellar mass in \citet{2004MNRAS.349..735B}, we obtain
$M_{\rm c,max}=0.28M_{\rm g}^{0.74}$.  Using this empirical relation
we estimate the mass of the most massive star cluster in the Milky Way
(MW) that formed from the most massive GMC.  According to our
analysis, the most massive GMC in the Milky Way is about $\sim
10^7M_{\odot}$ \citep{2011ApJ...729..133M}, and it could form a star
cluster with a mass of $\sim 3\times 10^4M_{\odot}$.  This estimate
for the most massive star cluster is consistent with that of
Westerlund 1 ($3\times 10^4M_{\odot}$), RSGC01, 02, 03 (3--$4\times
10^4M_{\odot}$), and Arches \citep[$2\times
  10^4M_{\odot}$][]{2010ARA&A..48..431P}.  For
clarity, we present in Figure \ref{fig:m_max} the observed MW cluster
and GMC.

When we adopt the relation $M_{\rm cut}=M_{\rm max, c}=0.20M_{\rm
  g}^{0.76}$, equation (\ref{eq:fit}) still fits satisfactorily to the
simulations, which then results in $\beta = -1.73\pm 0.17$ and
$A=0.64\pm0.29$.  And adopting $M_{\rm cut}=M_{\rm max, c}=6.3M_{\rm
  g}^{0.51}$ results in $\beta = -1.75\pm 0.17$ and $A=0.56\pm0.32$.
The values of $A$ and $\beta$ for each model are presented in Table
\ref{tb:results}.  We ignore models in with fewer than 5 clumps were
detected, because the resulting statistics becomes unreliable.  This
fit is presented as the dashed lines in Figures \ref{fig:CMF_sim}
(indicated with ``fitted model'').  This relation is slightly
shallower than the observed power-law ($\beta\simeq -2$) mass function
for massive clusters in Galactic disk and starburst galaxies
\citep[][and references therein]{2010ARA&A..48..431P}. We discuss this
in section 4. Note that the run-to-run variation in our simulations is
relatively large. The mass of the most massive cluster, for example,
varies by about an order of magnitude. Model m40k-d100-30K models are
an extreme case (see blue squares in Figure \ref{fig:CMF_sim}).

We also compared our simulation results with the observed
young-cluster mass-function in the MW \citep[data
  from][]{2003ARA&A..41...57L,2008A&A...487..557P,2010ARA&A..48..431P},
and in the Carina region \citep[data from][]{2011ApJS..194....9F},
which fits $\beta = -1.66 \pm 0.01$. 
This curve is presented in Figure\,\ref{fig:CMF_sim}
as the solid red curve.  The observed mass function is consistent with
our models for $M_{\rm g}\simeq 10^6 M_{\odot}$.

In our simulations the star formation efficiency and consequently the
maximum mass of a star clusters, depends slightly on the initial
density of the molecular cloud. The total stellar mass correlates with
$\sqrt{\rho_{\rm g}}$, because we assumed the local star formation
efficiency to depend on the local gas density. As a consequence the
star formation efficiency is highest in the densest regions.

\subsection{The Secular Evolution of the Cluster Mass Function due to Mergers}

In our simulations, clusters form hierarchically; more massive
clusters form from repeated mergers.  Gravitationally bound small
stellar clumps form quickly after the residual gas is removed. Each
clump corresponds to one of the density peaks in the parental gas
distribution (see Figure \ref{fig:gas_sd}). In the first few Myr, the
clumps grow in mass primarily by accreting smaller structures, but by
an age of about 2\,Myr, the cluster population has almost established
itself.  At that moment $\sim$ 30\% of the stars belong to a cluster
(see $M_{\rm s, cl}/M_{\rm s}$ in table \ref{tb:results}), and the
number of clusters drops in 10\,Myr to $\sim$ 80\%.

The most massive clusters tend to accrete some smaller clusters.  For
example in model m400k-d100-s1-30K, the number of clumps drops from 15
at 2\,Myr to 9 at an age of 10\,Myr.  During this phase the distance
between clusters increases because initially the entire system is
unbound, and the merger process stops in due time. In addition, the
smallest clusters that are still around at 2 Myr have disappeared by
an age of 10 Myr due to evaporation by relaxation process.

\begin{figure}
\begin{center}
\includegraphics[width=70mm]{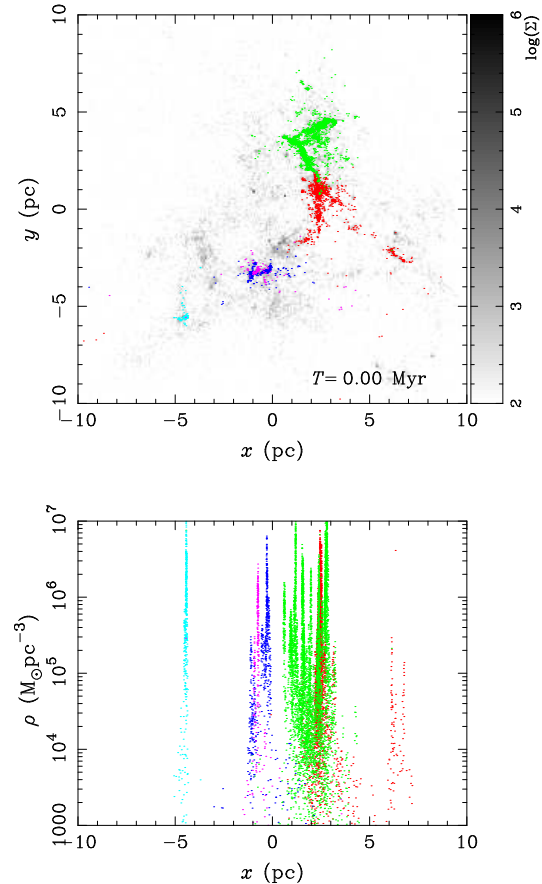}
\end{center}
\caption{Projected position of stars after the residual gas has been
  removed.  Each color identifies the cluster to which the star
  belongs at an age of 10\,Myr. The data is from model m400k-d100-s1-30K.  We
  used the same colors as in Figure \ref{fig:clump_finding}.  The projected
  stellar density is presented as a gray scale.  In the bottom
  panel we show the density distribution along one dimension, and gives
  the same data as is presented in Figure \ref{fig:gas_sd}, but for
  SPH particles which are converted to stars. The colors are the same
  as the top panel. Red, green, blue, cyan, and magenta clusters are relatively
  massive, and they are 2800, 5300, 690, 360, and 230 $M_{\odot}$,
  respectively.
  \label{fig:initial_pos}}
\end{figure}

We now investigate the location where the member stars of the clusters
formed.  In Figure \ref{fig:initial_pos} we have colored the initial
position of stars that at an age of 10 Myr belong to one
cluster. Stars located near each other within a few pc typically merge
to one cluster. The density peaks that at 10\,Myr still belong to a
cluster are presented in the bottom panel of Figure
\ref{fig:initial_pos}.  We identify several density peaks, each of
which eventually (at an age of 10\,Myr) corresponds to one
cluster. From the figure we see that a minimum gas density of $\sim
10^4M_{\odot}{\rm pc}^{-3}$ is required to form a cluster which
survives for 10 Myr.  This seems to be related the SFE-law we adopted
(see equation (\ref{eq:eff})). With a gas density of $1.6\times
10^4M_{\odot}{\rm pc}^{-3}$, the SFE $\sim 0.5$, and the mass in
regions with a density $\apgt 10^4M_{\odot}{\rm pc}^{-3}$ are
dominated by stars; as a consequence these regions easily survive the
gas expulsion.

The relation between the maximum cluster mass and molecular cloud mass
does not change appreciable from 2\,Myr to 10\,Myr.  In Figure
\ref{fig:m_max} we plot the results obtained from our simulations and
lines for $M_{\rm c, max}=6.3M_{\rm g}^{0.51}$ and 
$M_{\rm c, max}=0.20M_{\rm g}^{0.76}$.

\begin{figure}
\begin{center}
\includegraphics[width=70mm]{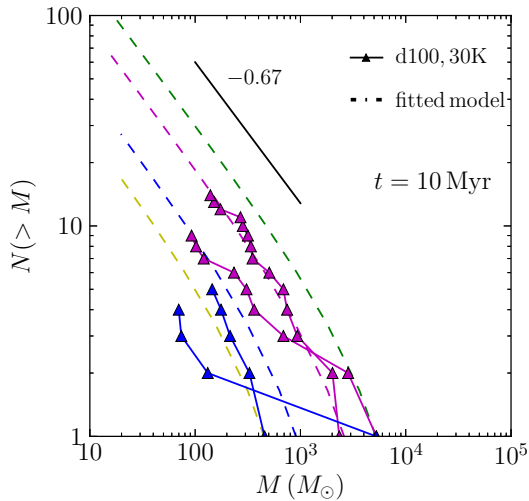}
\end{center}
\caption{Same as Figure \ref{fig:CMF_sim}, but for $t=10$ Myr. 
We adopt $A=0.63$, $\beta=-1.67$, and $M_{\rm c, max}=0.20M_{\rm g}^{0.76}$
for the fitting MF (see equation (\ref{eq:fit})).
\label{fig:CMF_sim_10Myr}}
\end{figure}

In Fig.\,\ref{fig:CMF_sim_10Myr} we present the mass distribution of
star clusters as a function of time.  The slope of the cluster mass
function becomes shallower between 2\,Myr and 10\,Myr, even though the
smaller number of clusters at later age.  At $t=10$ Myr the slope of
the cluster MF becomes $\beta=-1.56\pm 0.14$ for the simple fitting to
equation (\ref{eq:cum}).  If we assume that $M_{\rm c}=M_{\rm c, max}$
obtained from the simulations, we obtain $\beta=-1.53\pm 0.16$.
Assuming that $M_{\rm c, max}=0.20M_{\rm g}^{0.76}$, the slope of the
cluster MF becomes $\beta = -1.67 \pm 0.30$.  The best fit parameters
for each model are summarized in Table \ref{tb:results_10Myr}.  The
flatter slope is a natural consequence of the evaporation of the
smallest clusters and hierarchical merging, in which lower mass clumps
merge to more massive clumps.  After 10 Myr clusters stop merging,
because the entire system is unbound (see virial ratio of the system
in Table \ref{tb:models_nbody}).

\section{Total Cluster Mass Function in Disk Galaxies}

\begin{figure}
\begin{center}
\includegraphics[width=70mm]{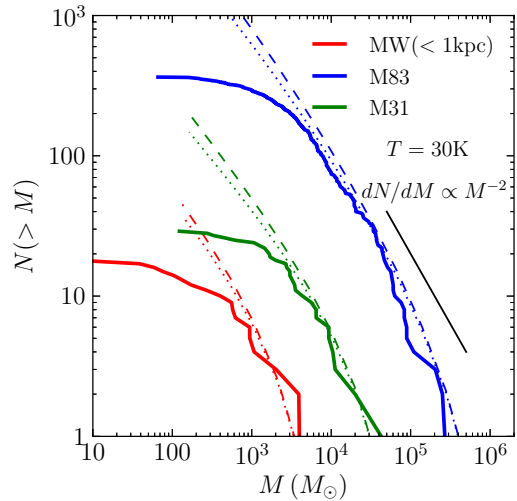}
\end{center}
\caption{Thick curves indicate the cumulative cluster MF for
  MW($D<1$kpc, $<30$ Myr), M31($<25$Myr), and M83 ($<70$Myr). Dashed
  and dotted curves indicate our model given by equation (\ref{eq:fit}) at 2
  Myr and at 10 Myr, respectively. 
  Here we adopt $\beta=-1.73$ and $A=0.64$ for $t=2$ Myr,
  $\beta=-1.67$ and $A=0.63$ for $t=10$ Myr, and 
  $M_{\rm c, max}=0.20M_{\rm g}^{0.76}$ for both.
  The observational data is from
  \citet{2011yCat..74179006B,2011MNRAS.417L...6B,2012MNRAS.419.2606B} 
  for M83 and \citet{2009ApJ...703.1872V} for M31.
\label{fig:cluster_MF_gal}}
\end{figure}

In order to compare our simulations with the observed cluster
mass-function in a galactic environment, we assume the GMC
mass-function to follow a power-law down to $100M_{\odot}$. We ignore
less massive clouds because they are unable to form sufficiently
massive clusters to compare with the observations.  For the Milky Way
we adopt a power of $-1.45$ \citep{2011A&A...536A..23P} and for M31 we
adopt $-0.9$ \citep{2013arXiv1306.2913K}. For M82 we also adopt a
power of $-1.45$, because there is insufficient data to properly fit
the distribution.  We assume that each individual GMC forms a
conglomerate of clusters that follow equation (\ref{eq:fit}).  Here we
adopt $\beta=-1.73$ and $A=0.64$ for $t=2$ Myr, $\beta=-1.67$ and
$A=0.63$ for $t=10$ Myr, and $M_{\rm c, max}=0.20M_{\rm g}^{0.76}$ for
both. As a consequence a distribution of GMCs forms a superposition of
multiple cluster mass-functions; one for each GMC in the galaxy.  The
mass functions for all young clusters in such a galaxy is presented in
Figure \ref{fig:cluster_MF_gal}, which we derived using the fitting
functions at 2\,Myr (dashes) and at 10 Myr (dotted curves).  We assume
that the total mass of the molecular gas is $2\times 10^7M_{\odot}$
and $8\times 10^8M_{\odot}$ for M31 and M83, respectively.  In this
calculation we do not allow any GMC to exceed half the total gas mass
in the galaxy. Observationally, the total mass of the molecular gas in
each galaxy is $3.6\times 10^8M_{\odot}$ for M31
\citep{2006A&A...453..459N} and $2.5\times 10^9M_{\odot}$ for M83
\citep{2002AJ....123.1892C}.  For young clusters in the MW and within
1 kpc of the Sun, we derive the total mass in molecular clouds of
$10^6M_{\odot}$ in order to match the cluster mass function. This
indicates that the total mass in molecular clouds within 1 kpc 
is $\sim 10^6M_{\odot}$. 

Because of the cut-off in the Schechter mass function, the power-law
slope of the superposed mass function is $\aplt -2$. This value is
consistent with cluster mass functions observed in nearby galaxies
\citep{2010ARA&A..48..431P}.  The solid curves in
Figure\,\ref{fig:cluster_MF_gal} indicate the observed MF for young
clusters in the MW (within a distance of $D=1$kpc from the Sun), M31,
and M83. With ``young'' we here indicate cluster with an age
comparable to the typical free-fall time of the GMC in each galaxy. If
we consider that GMCs collapse on this timescale and form stars, the
free-fall time scale of GMCs would be associated with the
star-formation time scale, as was suggested in
\citet{2012ApJ...745...69K}. The typical free-fall time scale
estimated from observations is $\sim 25$ Myr for M31 and $\sim
70$\,Myr for M83 \citep{2012ApJ...745...69K}.  These values are much
longer than the free-fall time scales in our models, but we consider
that the final SFE does not change much even if the initial free-fall
time scale is longer.  For the MW, we adopted a typical value for disk
galaxies of 30 Myr. Our model for the total cluster mass function
agrees with the observed cluster mass function in the MW, M31, and
M83.

We have not considered the evolution of star clusters beyond 10 Myr.
At these later ages the cluster-disruption process is expected to
continue to change the shape of the cluster mass function
\citep{2012MNRAS.419.2606B}, in particular by stellar mass loss,
internal two-body relaxation and the external influences of the
galactic tidal field. Some of the discrepancies between our
simulations and the observation are probably caused by our adopted
limited time scale of 10\,Myr, and by ignoring the global potential of
the parent galaxy.

\section{Discussion}

\subsection{The Self-Similarity between the Stellar Mass Function and the Cluster Mass Function}

The similarity between the cluster mass-function and the stellar
mass-function seems to originates from the self-similar structure of
GMC, as was also discussed in \citet{2002ApJ...577..206E}.  From an
observational perspective, the relation between the self-similar
structure of GMCs and star forming regions has been suggested by
\citep{1996ApJ...471..816E, 2004ARA&A..42..211E,2010ApJ...720..541S}.
As can be seen in Figure \ref{fig:initial_pos}, more massive clusters
tend to form in densiter regions. If the resolution of our
simulation was sufficiently high to resolve the formation of
individual stars, we expect to see spikes in the density distribution
which correspond to individual stars.  We then would expect that more
massive stars form in more massive and denser regions.  The most
massive star will then be born in the most massive cluster in the
system. The birth of the most massive star cluster as well as the 
most massive individual star is then limited by the highest mass and 
densest GMCs in the MW.  A similar argument was discussed in
\citet{2005ApJ...625..754W} and \citet{2010MNRAS.401..275W}.  

In our simulations we obtain the relation between GMC mass and the
maximum cluster mass of $M_{\rm c,max}=6.3M_{\rm g}^{0.51}$ or 
$M_{\rm c,max}=0.20M_{\rm g}^{0.76}$.  If we
apply these relations to the Milky Way, in which the most massive GMC is
$10^7M_{\odot}$ \citep{2011ApJ...729..133M}, it can form at most a
cluster of a few $10^4M_{\odot}$, which again forms a single star of at most
$\sim 100M_{\odot}$.  Note that even if the most massive stars are
born in the regions with the highest density, this location is not
necessarily associated with the most massive cluster. In this sense,
we consider that the $m_{\rm max}$-$M_{\rm c}$ relation is a
statistical result, as was suggested earlier by
\citet{2006ApJ...648..572E}, \citet{2010ARA&A..48..339B}, and
\citet{2010MNRAS.409L..54B}.

\begin{figure*}
\begin{center}
\includegraphics[width=130mm]{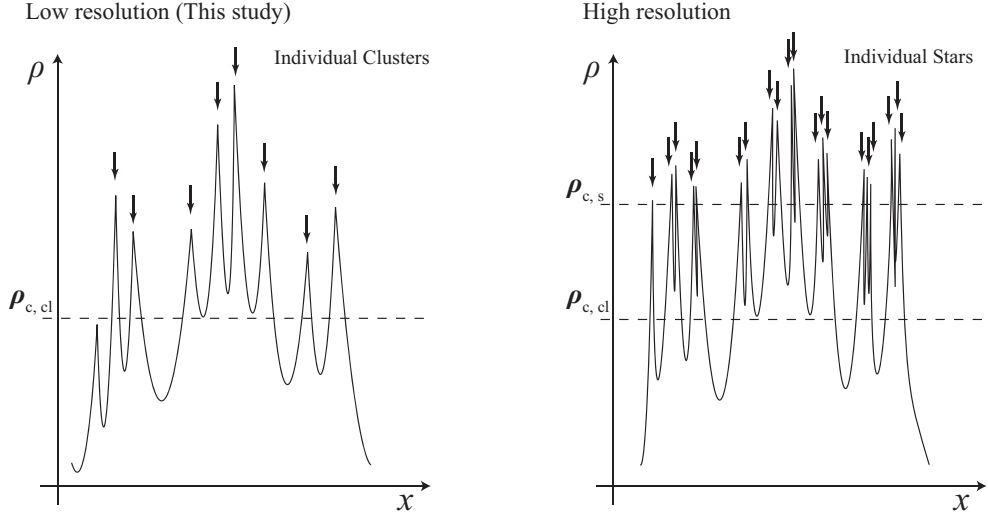}
\end{center}
\caption{Schematic representation of the density distribution of the
  formation process of cluster in relatively low-resolution
  simulations (left panel) and in an environment where the formation
  of individual stars is resolved (right).
\label{fig:rho_schematic}}
\end{figure*}

Although we observe a similarity between the largest structures in the
system; stellar mass function and cluster mass function, the slopes
are somewhat different. In order to follow the formation of individual
stars, higher resolution simulations of more massive clusters with the
appropriate physics would be required.  upon performing such
simulations we anticipate that the mass function of individual stars
would have a slope consistent with the Salpeter
\citep{1955ApJ...121..161S} slope, as was suggested by
\citep{2003MNRAS.343..413B,2003MNRAS.339..577B}.  In Figure
\ref{fig:rho_schematic}, we present a schematic picture of the
formation of a cluster (left) down to individual stars (right), as we
perceive it from our simulations.  In a simulations where individual
stars remain unresolved clusters can form, but by increasing the
resolution more fine-structure in the density distribution of the gas
will appear. Those density peaks are associated with individual stars,
and groups of peaks are associated with clusters of stars.  If the
peak density exceeds the critical density for star cluster formation
($\rho_{\rm c, cl}\sim 10^4M_{\odot}\mathrm{pc}^{3}$ in our
simulation), the cluster survives for at least 10 Myr. Some clusters
will merge to a larger and more massive cluster. If we were able to
resolve individual star formation, we would resolve the cluster
formation peaks to ensembles of peaks that correspond to individual
stars.

\subsection{Higher Temperature Simulations}
\begin{figure}
\begin{center}
\includegraphics[width=70mm]{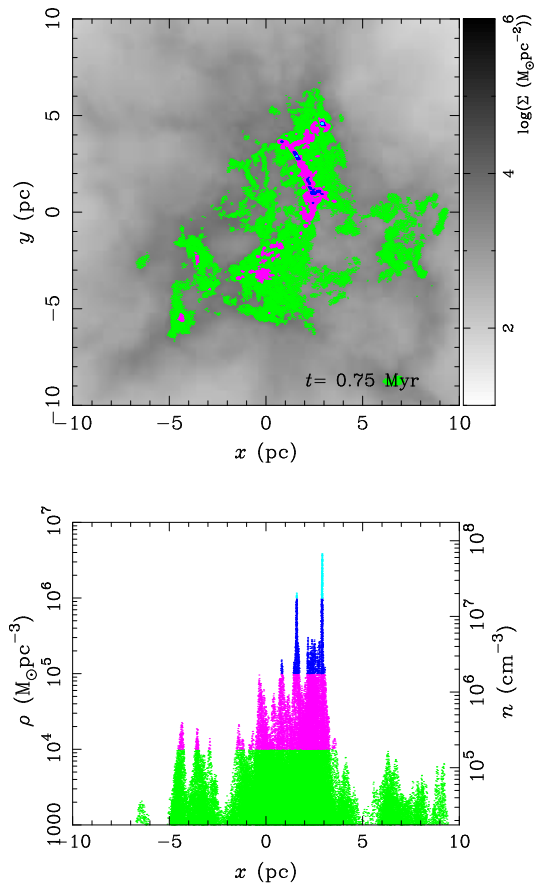}
\end{center}
\caption{Gas surface density at $0.9t_{\rm ff}$ for model
  m400k-d100-190K (top panel in gray scale).  Colors indicate dense
  region; cyan, blue, magenta, green indicates regions with the volume
  density of $>10^6$, $10^{5-6}$, $10^{4-5}$, and $10^{3-4}
  M_{\odot}$pc$^{-3}$, respectively. Gas volume density distribution
  projected on $x$-axis (bottom).  Colors are the same as in top
  panel.
  \label{fig:gas_sd_100K}}
\end{figure}

We adopt a temperature of 30K for our standard models, but we also
perform a series of simulations with a thermal energy of 1\% of the
kinetic energy in order to see the effect of the temperature. With
this setup, the gas temperature exceed 100 K. The other parameters are
the same as models with 30K. The initial conditions of the high
temperature models are summarized in Table \ref{tb:models_hydro_HT}.
The structures obtained from the hydro-simulation are considerably
smoother because the fragmentation scale is larger at a higher gas
temperature. In Figure \ref{fig:gas_star_100K} we present the gas
surface density at the end of the hydro-simulation for model
m400k-d100-190K, which is almost indistinguishable from
m400k-d100-30K, but which has a much higher gas temperature of
190K. The density peaks shown in Figure \ref{fig:gas_sd_100K} are less
pronounced than in the 30K models (see Figure \ref{fig:gas_sd}).
Although the number of small clumps seen as density peaks in Figure
\ref{fig:gas_sd} are smaller in the high-temperature simulations, the
eventual stellar distributions at 2 or 10 Myr, such as the shape of
cluster mass function obtained from the $N$-body simulations was not
much affected by the temperature.  In Table \ref{tb:models_nbody_HT}
we summarize the initial conditions for $N$-body simulations.

\begin{table*}
\begin{center}
\caption{Models\label{tb:models_hydro_HT} for hydrodynamical simulations with higher temperatures ('s' indicates the random seeds for the turbulence).}
\begin{tabular}{lccccccc}\hline \hline

Model  & Total mass & $N$ of particles & Radius  & Density & Thermal Energy & Temperature\\
      & $M_{\rm g}(M_{\odot})$ & $N_{\rm g}$ &$r_{\rm g}$(pc)  & $\rho_{\rm g}({\rm cm}^{-3})$ & $E_{\rm t}/E_{\rm k}$ & $T$ (K) \\ \hline
m5M-d10-s1-490K    & $5\times 10^6$ &$5\times 10^6$ & 49.0  & 170   & 0.01 & 490   \\
m1M-d100-s1-360K   & $1\times 10^6$ & $1\times 10^6$& 13.4  & $1.7\times10^3$  & 0.01 & 360   \\
m1M-d100-s6-360K   & $1\times 10^6$ & $1\times 10^6$& 13.4  & $1.7\times10^3$  & 0.01 & 360 \\
m1M-d100-s7-360K   & $1\times 10^6$ & $1\times 10^6$& 13.4  & $1.7\times10^3$  & 0.01 & 360  \\
m1M-d10-s1-170K    & $1\times 10^6$ & $1\times 10^6$ & 28.5  & 170  & 0.01 & 170  \\
m1M-d10-s4-170K    & $1\times 10^6$ & $1\times 10^6$ & 28.5  & 170  & 0.01 & 170 \\
m400k-d100-s1-190K & $4\times 10^5$ &$4\times 10^5$ & 10.0  & $1.7\times10^3$  & 0.01 & 190 \\
m400k-d100-s2-190K & $4\times 10^5$ &$4\times 10^5$ & 10.0 & $1.7\times10^3$  & 0.01 & 190\\
m400k-d100-s5-190K & $4\times 10^5$ & $4\times 10^5$ & 10.0  &$1.7\times10^3$  & 0.01 & 190  \\
m400k-d10-s1-90K   & $4\times 10^5$ &$4\times 10^5$ & 21.0 & 170  & 0.01 & 92  \\
m100k-d100-s1-80K  & $1\times 10^5$ & $1\times 10^5$ & 6.2 & $1.7\times10^3$  & 0.01 & 78\\
m40k-d100-s1-40K   & $4\times 10^4$ & $4\times 10^4$& 4.5 & $1.7\times10^3$  & 0.01 & 43  \\
\hline
\end{tabular}
\end{center}
\medskip
\end{table*}

\begin{table*}
\begin{center}
\caption{Models for $N$-body simulations based on higher temperature molecular clouds\label{tb:models_nbody_HT}}
\begin{tabular}{lccccc}\hline \hline

Model  & Total mass & $N$ of particles  & Virial ratio & SFE (Global) & SFE (Dense)\\
     & $M_{\rm s}(M_{\odot})$ & $N_{\rm s}$  & $Q_{\rm vir}$  & $\epsilon$ & $\epsilon_{\rm d}$\\ \hline
m5M-d10-s1-490K      & $1.6\times 10^5$ & 155972  & 5.4   & 0.032 & 0.25\\
m1M-d100-s1-360K  & $5.8\times 10^4$ & 57642 & 3.0  & 0.058 & 0.20 \\
m1M-d100-s6-360K  & $4.1\times 10^4$ & 40510 & 7.6  & 0.041 & 0.13 \\
m1M-d100-s7-360K  & $6.9\times 10^4$ & 68901 & 1.0  & 0.070 & 0.19\\
m1M-d10-s1-170K   & $3.1\times 10^4$ & 31023 &  4.6   & 0.032 & 0.18 \\
m1M-d10-s4-170K   & $3.6\times 10^4$ & 36224 &  3.7   & 0.037 & 0.30  \\
m400k-d100-s1-190K & $3.2\times 10^4$ & 31868 &  2.8  & 0.078 & 0.26 \\
m400k-d100-s2-190K & $3.0\times 10^4$ & 30496 &  1.5  & 0.075 & 0.22\\
m400k-d100-s5-190K & $2.5\times 10^4$ & 25419 &  2.8  & 0.062 & 0.19 \\
m400k-d10-s1-90K   & $1.5\times 10^4$ & 15124 & 3.2   & 0.037 & 0.35 \\
m100k-d100-s1-80K  & $6.4\times 10^3$ & 6474  & 1.4  & 0.063 & 0.20\\
m40k-d100-s1-40K   & $2.5\times 10^3$ & 2498  & 2.0  & 0.062 & 0.19\\
\hline
\end{tabular}
\medskip

\end{center}
\end{table*}

\begin{figure}
\begin{center}
\includegraphics[width=\columnwidth]{f11.eps}
\end{center}
\caption{Gas surface density at an age of $0.9t_{\rm ff}$ for model
  m400k-d100-190K. \label{fig:gas_star_100K}}
\end{figure}

\begin{figure*}
\begin{center}
\includegraphics[width=0.6\columnwidth]{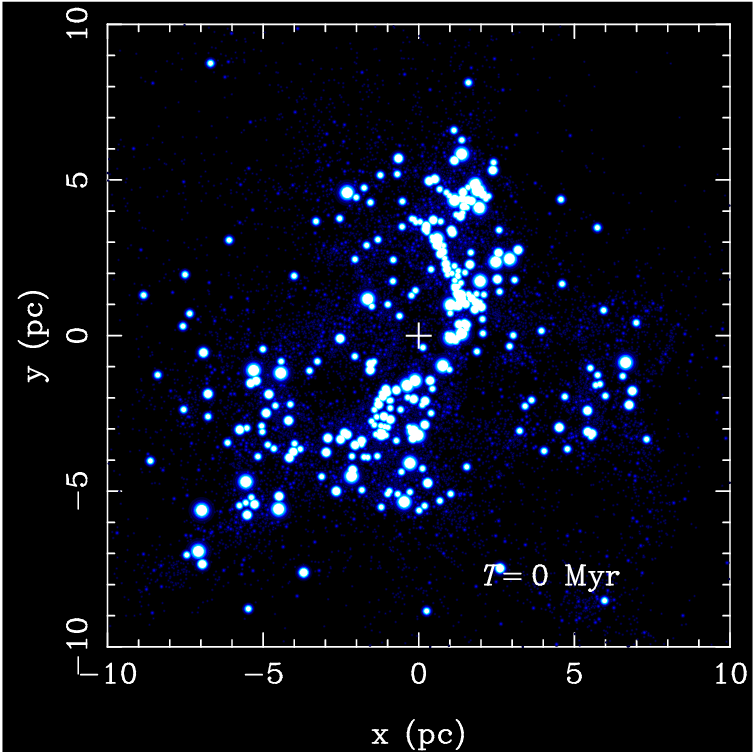}
\includegraphics[width=0.6\columnwidth]{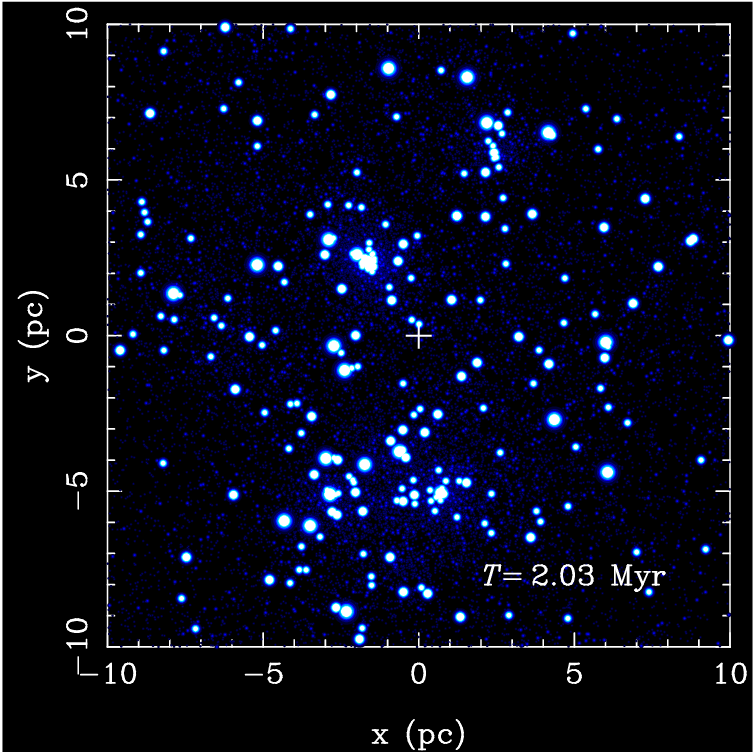}
\includegraphics[width=0.6\columnwidth]{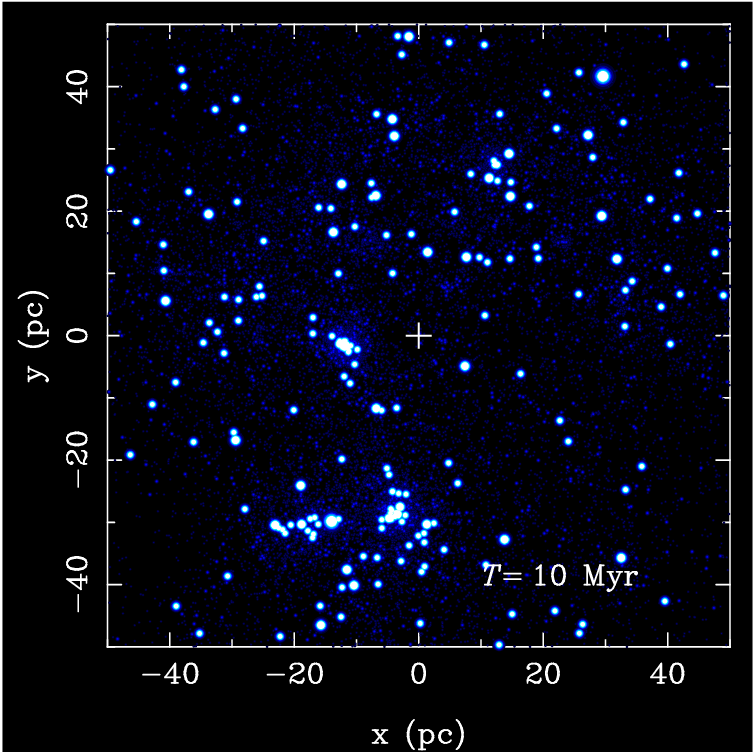}
\end{center}
\caption{The snapshots of the $N$-body simulation for model 
 m400k-d100-190K.
\label{fig:stars_100K}}
\end{figure*}

The evolution of the stellar system for model m400k-d100-190K is shown
in Figure \ref{fig:stars_100K}. The position and mass of the clusters
formed at 10 Myr in m400k-d100-190K are different that those resulting
from model m400k-d100-30K. The stellar composition of the most massive
clusters and their initial locations in space is hardly affected by
the different temperatures. In Figure \ref{fig:initial_pos_100K} we
present, for model m400k-d100-190K, the stellar positions that are
contained in the five most massive clusters at an age of 10 Myr.

\begin{figure}
\begin{center}
\includegraphics[width=80mm]{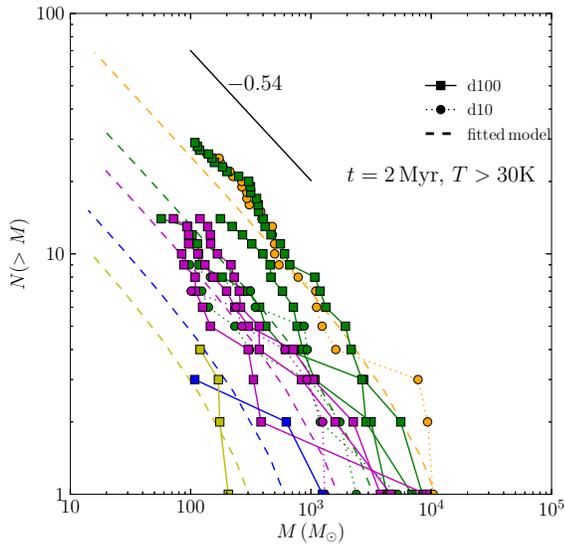}
\end{center}
\caption{Same as Figure \ref{fig:CMF_sim}, but for models with $>30$K. 
  The colors represent
  different masses of initial molecular clouds: orange, green, magenta,
  blue, and yellow indicate $M_{\rm g}$ of $5\times 10^6, 10^6,
  4\times 10^5, 10^5$ and $4\times 10^4 M_{\odot}$, respectively.  The dashed
  curves give the fitted MF (see equation (\ref{eq:fit})).
  We adopt $A=0.60$, $\beta=-1.54$, and $M_{\rm c, max}=0.20M_{\rm g}^{0.76}$. 
  \label{fig:CMF_sim_warm}}
\end{figure}

\begin{figure}
\begin{center}
\includegraphics[width=80mm]{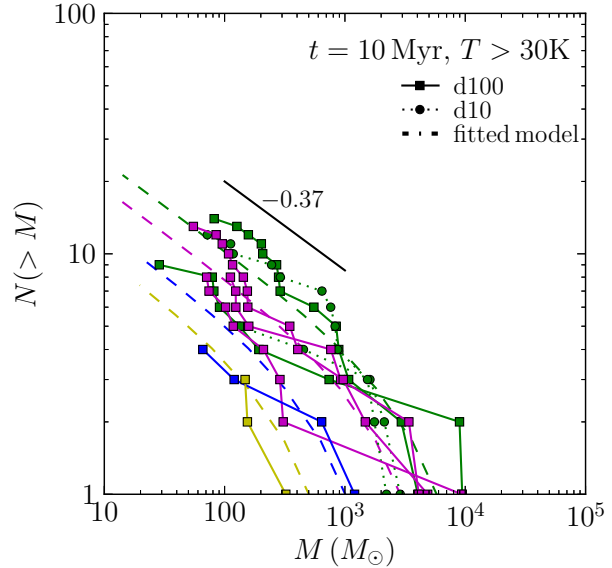}
\end{center}
\caption{Same as Figure \ref{fig:CMF_sim_10Myr}, but for models with $>30$K. 
   The dashed
  curves give the fitted MF (see equation (\ref{eq:fit})).
  We adopt $A=0.78$, $\beta=-1.37$, and $M_{\rm c, max}=0.20M_{\rm g}^{0.76}$.
  \label{fig:CMF_sim_warm_10Myr}}
\end{figure}

\begin{figure}
\begin{center}
\includegraphics[width=70mm]{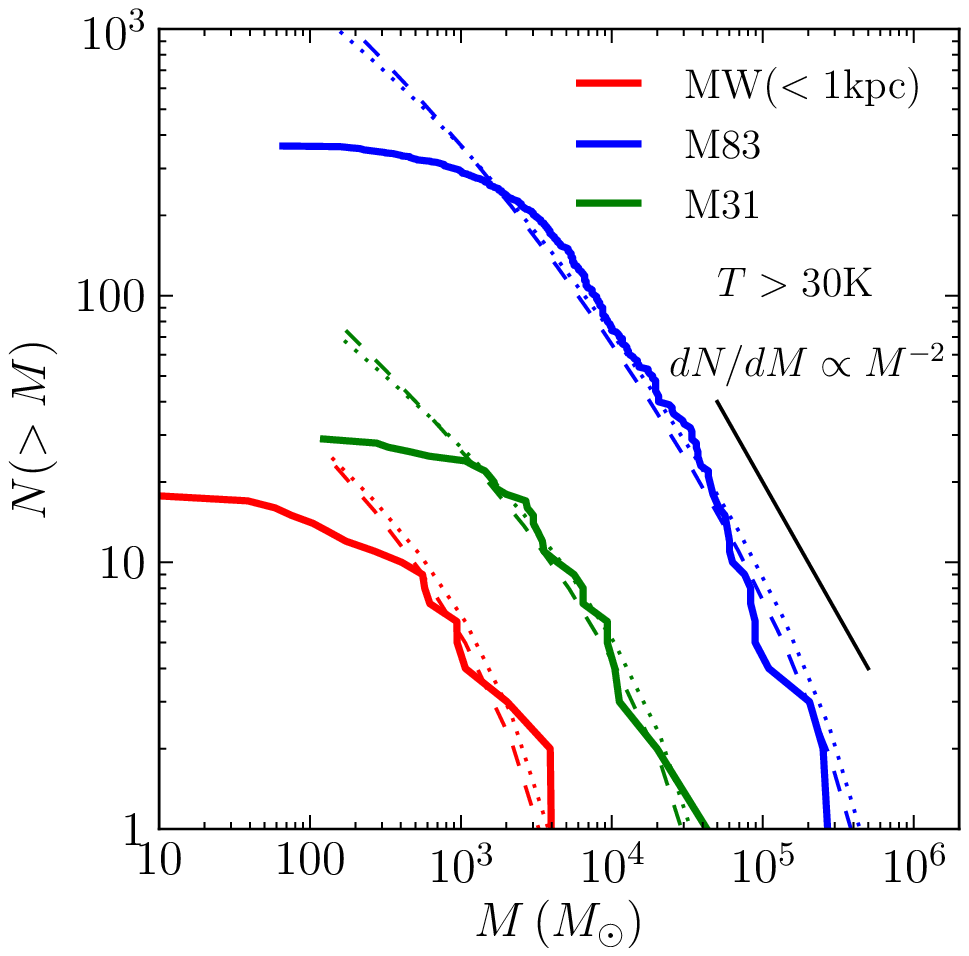}
\end{center}
\caption{Same as Figure \ref{fig:cluster_MF_gal} but for 
high temperature models. Here we adopt $\beta=-1.52$ and $A=0.60$ for $t=2$ Myr
(dashed lines) and $\beta=-1.37$ and $A=0.78$ for $t=10$ Myr (dotted lines),
respectively, and adopt $M_{\rm c, max}=0.20M_{\rm g}^{0.76}$ for both.
\label{fig:cluster_MF_gal_warm}}
\end{figure}

We investigated the cluster MF for these models at $t=2$ and 10 Myr.
The resulting cluster MFs are shown in Figures \ref{fig:CMF_sim_warm}
and \ref{fig:CMF_sim_warm_10Myr} for $t=2$ and 10 Myr, respectively.
We also fit the results to equation (\ref{eq:fit}).  The fitted
$\beta$ and $A$ for each model are summarized in Tables
\ref{tb:results_HT} and \ref{tb:results_10Myr_HT}.  When we adopt the
actual values of the most massive clusters as $M_{\rm c, max}$ we
obtain $\beta = -1.52\pm0.21$ for the high temperature models, but
$\beta=-1.62\pm0.22$ when we include the 30K models. The mass of the
most massive clusters does not much different from those for 30K
models.  The relation between the initial gas mass and the mass of the
most massive cluster is $M_{\rm c, max}=26M_{\rm g}^{0.39}$ at $t=2$
Myr only for high temperature models.  By adding artificial points at
$M_{\rm c, max}=0.01M_{\odot}$ and $M_{\rm g}=0.02M_{\odot}$, we
obtain $M_{\rm c, max}=0.17M_{\rm g}^{0.73}$ at $t=2$ Myr.  If we
assume $M_{\rm c, max}=0.20M_{\rm g}^{0.76}$ (the same as that
obtained in 30K models), we obtain the averaged power-law slope of the
fitting function of $\beta=-1.52\pm0.22$ and $\beta= -1.37\pm0.20$ at
$t=2$ and 10 Myr, respectively. These slopes are slightly shallower
than those in the low temperature simulations, but both values are
consistent within the uncertainty. We also obtain $A=0.60\pm0.29$ and
$A=0.78\pm0.34$ at $t=2$ and 10 Myr.  With these values, the total
cluster MF for galaxies obtained from our model is consistent with the
observations (see Figure \ref{fig:cluster_MF_gal_warm}).

\begin{table*}
\begin{center}
\caption{The results of simulations at $t=2$Myr\label{tb:results_HT}.
  For $\beta_2$ and $A$, we assume $M_{\rm c, max}=0.20M_{\rm
    g}^{0.76}$.  Averaging the results, we obtain $\beta_1=-1.52\pm
  0.21$, $\beta_2 = -1.54\pm 0.16$, and $A=0.60\pm 0.29$ only for high
  temperature models and $\beta_1=-1.62\pm 0.22$, $\beta_2 = -1.64\pm
  0.19$, and $A=0.62\pm 0.28$ for all models including 30K models.}
\begin{tabular}{lcccccc}\hline \hline
Model & $ M_{\rm s, cl}/M_{\rm s}$ & $M_{\rm c, max}(M_{\odot})$ & $N_{\rm c}$ & $\beta_1$ & $\beta_2$ & $A$   \\ \hline
m5M-d10-s1-490K &  0.33 & $1.0\times 10^4$ & 27 & $-1.61\pm 0.03$  & $-1.63\pm 0.03$ & $0.37\pm 0.05$ \\ 
m1M-d100-s1-360K & 0.33 & $8.4\times 10^3$ & 14 & $-1.43\pm 0.06$  & $-1.43\pm 0.06$ & $0.65\pm 0.15$ \\ 
m1M-d100-s6-360K & 0.34 &  $4.3\times 10^3$ & 14 & $-1.58\pm 0.05$ & $-1.62\pm 0.05$ & $0.57\pm 0.09$\\ 
m1M-d100-s7-360K & 0.39 &  $7.0\times 10^3$ & 29 & $-1.52\pm 0.02$ & $-1.52\pm 0.02$ & $1.24\pm 0.10$\\ 
m1M-d10-s1-170K &  0.26 & $5.2\times 10^3$ & 7 & $-1.39\pm 0.06$  & $-1.41\pm 0.05$ & $0.46\pm 0.09$ \\ 
m1M-d10-s4-170K &  0.21 &  $2.4\times 10^3$ & 9 & $-1.23\pm 0.04$ & $-1.40\pm 0.08$ & $0.69\pm 0.16$\\ 
m400k-d100-s1-190K & 0.29 & $3.8\times 10^3$ & 14 & $-1.59\pm 0.05$ & $-1.45\pm 0.05$ & $0.76\pm 0.12$\\ 
m400k-d100-s2-190K & 0.35  & $9.1\times 10^3$ & 10 & $-1.78\pm 0.15$ & $-1.76\pm 0.15$ & $0.19\pm 0.09$ \\ 
m400k-d100-s5-190K  &  0.37 & $1.5\times 10^3$ & 14 & $-1.87\pm 0.07$ & $-1.85\pm 0.07$ & $0.29\pm 0.06$ \\ 
m400k-d10-s1-90K & 0.31 & $1.3\times 10^3$ & 7 & $-1.16\pm 0.06$ & $-1.35\pm 0.08$ & $0.81\pm 0.17$\\ 
m100k-d100-s1-80K & 0.31 &$1.3\times 10^3$ & 3 & - & - & -\\ 
m40k-d100-s1-40K &  0.28 & $2.1\times 10^2$ & 4 & - & - & - \\ 
\hline
\end{tabular}
\end{center}
\medskip
\end{table*}

\begin{table*}
\begin{center}
\caption{The results of simulations at
  10Myr\label{tb:results_10Myr_HT}.  Averaging the results, we obtain
  $\beta_1=-1.36\pm 0.20$, $\beta_2 = -1.37\pm 0.20$, and $A=0.78\pm
  0.34$ for high temperature models and $\beta_1=-1.43\pm 0.21$,
  $\beta_2 = -1.46\pm 0.27$, and $A=0.73\pm 0.34$ for all models
  including 30K models.  Here, we assume $M_{\rm c, max}=0.20M_{\rm
    g}^{0.76}$.}
\begin{tabular}{lccccc}\hline \hline
Model & $M_{\rm c, max}(M_{\odot})$ & $N_{\rm c}$ & $\beta_1$ & $\beta_2$ & $A$   \\ \hline
m1M-d100-s1-360K  & $9.4\times 10^3$ & 9 & $-1.31\pm 0.07$ & $-1.55\pm 0.04$ & $0.55\pm 0.08$\\ 
m1M-d100-s6-360K  & $4.0\times 10^3$ & 14 & $-1.43\pm 0.04$ & $-1.22\pm 0.07$& $0.79\pm 0.18$\\ 
m1M-d10-s1-170K &   $2.9\times 10^3$ & 5 & $-1.08\pm 0.05$ & $-1.37\pm 0.05$& $0.88\pm 0.17$\\ 
m1M-d10-s4-170K & $2.2\times 10^3$   & 12 & $-1.22\pm 0.03$ & $-1.07\pm 0.11$& $1.28\pm 0.49$\\ 
m400k-d100s1-190K &  $4.1\times 10^3$ & 8 & $-1.24\pm 0.08$& $-1.23\pm 0.08$ & $1.18\pm 0.30$\\ 
m400k-d100-s2-190K &  $9.3\times 10^3$ & 8 & $-1.70\pm 0.09$ & $=1.68\pm 0.17$ & $0.20\pm 0.11$\\ 
m400k-d100-s5-190K &  $4.9\times 10^3$ & 13 & $-1.41\pm 0.05$& $=1.49\pm 0.08$ & $0.61\pm 0.15$\\ 
m100k-d100-s1-80K & $1.2\times 10^3$  & 4 & - & - & - \\ 
m40k-d100-s1-40K & $3.2\times 10^2$   & 3 & - & - & - \\ 
\hline
\end{tabular}
\end{center}
\medskip
\end{table*}

The high temperature we assumed here ($>100$K) is not realistic for a
model of molecular clouds. The main clusters formed from initial
conditions with a high temperature, however, are not much different
from those formed from molecular cloud with a realistic
temperature. In low-temperature models, there are more clumps in a
region which finally merge to a cluster. The clumps merge within a
local dynamical time scale irrespective of the gas temperature.  The
clumpy structure is lost quickly in the merger process, mainly due to
violent relaxation. The formation of massive cluster is therefore not
affected by the unrealistically high temperature of the gas.  This
result is practical when applying this same method to a cluster
formation simulations on a galactic scale. In evironments where
temperature of $\apgt 100$\,K are reached, which is a quite typical
threshold temperature of star forming regions in galaxy simulations,
the temperature is sufficiently low to result in clustered star
formation.  In those cases we do not necessarily resolve the
low-temperature and high density regions.

\begin{figure}
\begin{center}
\includegraphics[width=70mm]{f14.eps}
\end{center}
\caption{Projected position of stars after the residual gas has been
  removed.  Each color identifies the cluster to which the star
  belongs at an age of 10\,Myr. The data is from model m400k-d100-190K.  We
  used the same colors as in Figure \ref{fig:clump_finding}.  The
  stellar surface density is presented as a gray scale.  In the bottom
  panel we show the density distribution along one dimension, and gives
  the same data as is presented in Figure \ref{fig:gas_sd}, but for
  SPH particles which are converted to stars. The colors are the same
  as the top panel. Red, green, blue, cyan, and magenta clusters are relatively
  massive, and they are 4100, 3400, 760, 920, and 123 $M_{\odot}$,
  respectively.
  \label{fig:initial_pos_100K}}
\end{figure}

\section{Summary}

We performed $N$-body simulations of ensembles of young star clusters.
The initial conditions of our dynamical simulations are obtained from
smoothed particles hydrodynamical simulations of turbulent molecular
clouds. Both calculations, the collapsing molecular cloud and the
gravitational dynamical simulations, are performed individually and
are separated. As a result, our calculations are not self consistent,
but there is a natural causality.

In our approach we start with the hydrodynamical simulation, which we
continue for about a free-fall time scale. We subsequently analyze the
distribution of the gas and assign stellar mass to individual SPH
particles. In this procedure, mass is locally not conserved, but
globally it is. Only those SPH particles that have a local density in
excess of the star formation threshold are assumed to form star.  This
procedure of adopting a simple density threshold results in a local
(core) star-formation efficiency of about 30 per-cent, but only about
1 per cent of the total mass in gas is converted to stars.  Individual
stars are assigned the positions and velocities of the SPH particles
from which they were generated. These mass, positions and velocities
are adopted as the initial conditions for the gravitational dynamics
simulations.

The distribution of stars resulting from this procedure show a strong
hierarchical structure. The distribution of the mass of the stellar
clumps are consistent with the Schechter function.  In the first
2\,Myr, after the stars formed, the mass function of clusters
resemblances a power-law with a slope of $\simeq -1.73$. This slope
becomes shallower with time because of hierarchical merging, to reach
a slope of $\simeq -1.67$ at an age of 10 Myr.

The shape of the cluster mass function in our simulations is
consistent with the one observed in active star forming regions in the
Milky Way, such as in the Carina region.  We find a relation between
the mass of the GMC and the most massive cluster: $6.3M_{\rm
  g}^{0.51}$ (or $0.20M_{\rm g}^{0.76}$ if we assume that this
relation continues down to the stellar mass scale). This relation is
consistent with the observed cluster distribtion in M51
\citep{2013ApJ...779...44H}, and it is similar to the
observed relation between the most massive star and the total mass of
the cluster \citep{2010MNRAS.401..275W}. We therefore conclude that
star-forming regions and star cluster-forming regions have a
self-similar structure down to the formation of individual stars.

Using our simulations we estimate the global galactic cluster MF in
the MW, M31, and M83. We satisfactory fitted the distribution of
cluster masses for each of these galaxies. The galactic cluster MFs
obtained from our model have a power of $\aplt -2$, 
which is consistent with observed cluster MF nearby galaxies.

\section*{Acknowledgments}
The authors would like to thank Annie Hughes for providing the 
observational data of M51, Nate Bastian for useful comments on 
the manuscript, and Inti Pelupessy, Kengo Tomida, and
Yusuke Tsukamoto for their advise on the simulations.  
The authors are also grateful to the anonymous referee for the 
detailed comments. This work was
supported by Postdoctoral Fellowship for Research Abroad of the Japan
Society for the Promotion of Science (JSPS) the Netherlands Research
Council NWO (grants \#612.071.305 [LGM], \#639.073.803 [VICI] and
\#614.061.608 [AMUSE]) and by the Netherlands Research School for
Astronomy (NOVA).  Numerical computations were carried out on Cray
XT-4 and XC30 CPU-cluster at the Center for Computational Astrophysics
(CfCA) of the National Astronomical Observatory of Japan and the
Little Green Machine (GPU cluster) at Leiden Observatory.

\bibliographystyle{mn}
\bibliography{reference}

\label{lastpage}

\end{document}